\documentclass[%
 reprint,
superscriptaddress,
preprintnumbers,
 amsmath,amssymb,
 aps,
 prx,
]{revtex4-2}
\usepackage[hyperfootnotes=false]{hyperref}
\hypersetup{
   colorlinks=true,
   linkcolor=blue,
   urlcolor=blue,
   linktoc=all,
   citecolor=blue
          }
\usepackage{graphicx}
\usepackage{dcolumn}
\usepackage{bm}
\usepackage{physics}

\usepackage{xcolor}
\usepackage{bbold}
\usepackage[normalem]{ulem}
\usepackage{mathtools}

\begin{document}

\preprint{
UMD-PP-024-07}

\title{Analog Quantum Simulator of a Quantum Field Theory \\with Fermion-Spin Systems in Silicon
}

\author{Ali~Rad}
\affiliation{Joint Center for Quantum Information and Computer Science, NIST/University of Maryland, College Park, Maryland 20742 USA.}
\affiliation{Joint Quantum Institute, NIST/University of Maryland, College Park, Maryland 20742 USA.}
\affiliation{The NSF Institute for Robust Quantum Simulation, University of Maryland, College Park, Maryland 20742, USA}
\author{Alexander~Schuckert}
\affiliation{Joint Center for Quantum Information and Computer Science, NIST/University of Maryland, College Park, Maryland 20742 USA.}
\affiliation{Joint Quantum Institute, NIST/University of Maryland, College Park, Maryland 20742 USA.}
\author{Eleanor~Crane}
\affiliation{Joint Center for Quantum Information and Computer Science, NIST/University of Maryland, College Park, Maryland 20742 USA.}
\affiliation{Joint Quantum Institute, NIST/University of Maryland, College Park, Maryland 20742 USA.}
\affiliation{Department of Electrical Engineering and Computer Science, Massachusetts Institute of Technology, Cambridge, Massachusetts 02139, USA}
\author{Gautam~Nambiar}
\affiliation{Joint Quantum Institute, NIST/University of Maryland, College Park, Maryland 20742 USA.}
\author{Fan~Fei}
\affiliation{Atom Scale Device Group, National Institute of Standards and Technology, Gaithersburg, MD 20899, USA.}
\author{Jonathan~Wyrick}
\affiliation{Atom Scale Device Group, National Institute of Standards and Technology, Gaithersburg, MD 20899, USA.}
\author{Richard~M.~Silver}
\affiliation{Atom Scale Device Group, National Institute of Standards and Technology, Gaithersburg, MD 20899, USA.}
\author{Mohammad~Hafezi}
\affiliation{Joint Center for Quantum Information and Computer Science, NIST/University of Maryland, College Park, Maryland 20742 USA.}
\affiliation{Joint Quantum Institute, NIST/University of Maryland, College Park, Maryland 20742 USA.}
\affiliation{The NSF Institute for Robust Quantum Simulation, University of Maryland, College Park, Maryland 20742, USA}
\author{Zohreh~Davoudi}
\email{davoudi@umd.edu}
\affiliation{Joint Center for Quantum Information and Computer Science, NIST/University of Maryland, College Park, Maryland 20742 USA.}
\affiliation{The NSF Institute for Robust Quantum Simulation, University of Maryland, College Park, Maryland 20742, USA}
\affiliation{Maryland Center for Fundamental Physics and Department of Physics,
University of Maryland, College Park, MD 20742 USA.}
\author{Michael~J.~Gullans}
\email{mgullans@umd.edu}
\affiliation{Joint Center for Quantum Information and Computer Science, NIST/University of Maryland, College Park, Maryland 20742 USA.}
\affiliation{The NSF Institute for Robust Quantum Simulation, University of Maryland, College Park, Maryland 20742, USA}
\affiliation{Atom Scale Device Group, National Institute of Standards and Technology, Gaithersburg, MD 20899, USA.}

\date{\today}

\begin{abstract}

Simulating fermions coupled to spin degrees of freedom, relevant for a range of quantum field theories, represents a promising application for quantum simulators. Mapping fermions to qubits is challenging in $2+1$ and higher spacetime dimensions, and mapping bosons demands substantial 
quantum-computational overhead. These features complicate the realization of mixed fermion-boson quantum systems in digital quantum computers.
We propose a native fermion--(large-)spin analog quantum simulator by utilizing dopant arrays in silicon. Specifically, we show how to use a dynamical lattice of coupled nuclear spins and conduction-band electrons to realize a quantum field theory: an extended Jackiw-Rebbi model involving coupled fermions and quantum rotors. We demonstrate the feasibility of observing dynamical mass generation and a confinement-deconfinement quantum phase transition in 1+1 dimensions on this platform, even in the presence of strong long-range Coulomb interactions.
Furthermore, we employ finite-temperature Hartree-Fock-Bogoliubov simulations to investigate the dynamics of mass generation in two-dimensional square and honeycomb arrays, showing that this phenomenon can be simulated with realistic experimental parameters. Our findings reveal two distinct phases, and demonstrate robustness against the addition of Coulomb interactions. Finally, we discuss experimental signatures of the phases through transport and local charge sensing in dopant arrays. This study lays the foundation for quantum simulations of quantum field theories exhibiting fermions coupled to spin degrees of freedom using donors in silicon.

\end{abstract}

\maketitle


\section{Introduction}
\begin{figure}[t]
\includegraphics[width=0.45\textwidth]{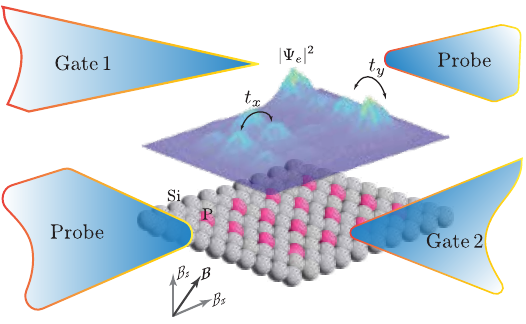}
\centering
\caption{\textbf{Donor array in silicon.} A schematic illustrating electron localization and tunneling behaviors in a silicon phosphorus-donor array. The lattice model approaches continuity as donor distance tends to zero and lattice size tends to infinity. Electrons can hop in both $x$ and $y$ directions with tunneling rates $t_x$ and $t_y$. An external magnetic field, $\bm{B}$, can be applied to polarize electrons in a given direction. The probes (charge or conductance), along with gates, allow indirect inference and control of the internal electron-density configuration, $|\Psi_e|^2$, within the central island, facilitating a comprehensive understanding of electron behavior and phases of the system in the donor lattice structure.
}\label{Fig:array_schematic}
\end{figure}
\noindent
Understanding the origin of the mass of elementary and composite particles has long driven the field of particle and hadronic physics. A spontaneous symmetry breaking of gauge symmetries via the Higgs mechanism gives rise to the mass of elementary particles in the Standard Model~\cite{Englert:1964et,Higgs:1964pj}, while spontaneous breaking of a chiral symmetry greatly influences the spectrum of hadrons in quantum chromodynamics (QCD)~\cite{PhysRev.122.345,PhysRev.124.246,pich1995chiral}. Despite this understanding, a complete picture of dynamical mass generation in QCD in, e.g., systems with nonzero fermion density and/or in out-of-equilibrium conditions, and the connection between chiral symmetry breaking and confinement-deconfinement phase transition, are yet to be developed~\cite{Kogut:2004su,Kogut:1982rt,Shuryak:2016vow,Brambilla:2014jmp}. While the method of lattice QCD with the aid of classical computing has opened the door to illuminate part of this picture~\cite{Gattringer:2010zz,FlavourLatticeAveragingGroupFLAG:2021npn,kronfeld2022lattice,Davoudi:2022bnl}, some questions remain intractable classically. Quantum simulation of quantum (gauge) field theories~\cite{banuls2020simulating,klco2021standard,bauer2023quantumprx,bauer2023quantumnature, di2023quantum, halimeh2023cold} provides a unique avenue to explore inherently quantum, nonperturbative, and dynamical phenomena such as quantum phase transitions and mass generation in QCD and other strongly-coupled quantum field theories. We aim to introduce a solid-state analog quantum simulator to explore such questions in a quantum field theory, setting the stage for future experimental implementations.

\begin{figure}[t!]
\includegraphics[width=\columnwidth]
{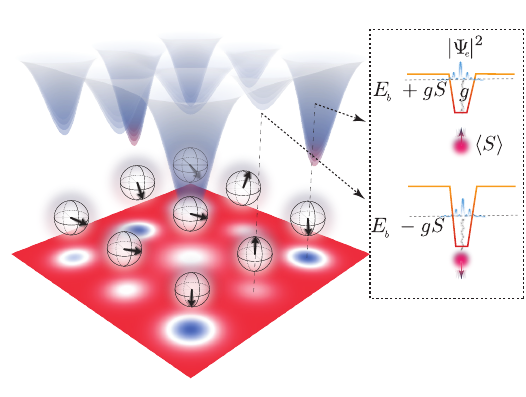}
\centering
\caption{\textbf{Schematic of a dynamic lattice.} 
Each individual donor nuclear-spin orientation (illustrated on the Bloch sphere) can dynamically change and locally alter the potential landscape at each fermionic donor-electron site,
acting as a dynamical mass term akin to a Jackiw-Rebbi rotor coupled to fermionic matter. Each spin, $\bm{S}$, which in our platform is represented by the nuclear spin of a donor, interacts with the local density of fermions, $|\Psi_e|^2$, via the hyperfine interaction with strength $g$. 
Depending on the orientation of the nuclear spin at each site, the energy level of the system at each site is in the range of $E_b \pm gS$, where $E_b$ represents the binding energy of the electron and nuclear spin.
\label{Fig:1}
}
\end{figure}

A range of models for mass generation have been proposed, such as the Nambu model and the Gross-Neveu model~\cite{PhysRev.122.345,PhysRev.124.246,hatsuda1994qcd,PhysRevD.10.3235,Gross:1975vu,zamolodchikov1978exact,RevModPhys.64.649}, allowing for qualitative understanding in systems that are simpler to study. An especially important case is \emph{dynamical} mass generation, where mass in the fermion field arises from the spontaneous breakdown of chiral symmetry~\cite{hatsuda1994qcd,witten1978chiral,zamolodchikov1978exact,berg1978exact}, which may be observed when the confinement-deconfinement phase transition is crossed. One of the simplest models which has been shown to capture the physics of dynamical mass generation and the confinement-deconfinement quantum phase transition is the Jackiw-Rebbi model~\cite{jackiw1976solitons, gonzalez2020rotor}. This model consists of matter particles interacting with a dynamical lattice, i.e., a lattice hosting dynamical bosons, see Fig.~\ref{Fig:1}. A first analog-quantum-simulation proposal for simulating the Jackiw-Rebbi model in (1+1)D was developed in Ref.~\cite{gonzalez2020rotor}, but higher dimensional versions were left unexplored. Digital quantum computation could simulate this model using a Jordan-Wigner mapping of fermions~\cite{Jordan1928,Kivlichan2018,Stanisic2022,hemery2023}, and e.g., a binary mapping of bosons~\cite{BOM}, to qubits, and by digitizing the time-evolution operator in terms of a universal set of gates.
These approaches, therefore, demand a large computational overhead and a high level of quantum control. By contrast, analog quantum simulations can produce faithful results, often by a natural mapping of degrees of freedom and by allowing for a continuous evolution. Therefore, finding an analog-simulation method using native fermionic and bosonic matter would greatly improve the scalability of the simulation of the Jackiw-Rebbi model and other quantum field theories in the near term.

We propose a (1+1)D and (2+1)D native fermionic-bosonic platform for quantum simulation of the rotor Jackiw-Rebbi model, based on donors in silicon (see Fig.~\ref{Fig:array_schematic}). The fermionic degrees of freedom are directly encoded in conduction-band electrons that populate the array, removing the necessity to map fermions to qubits. Dynamical lattices with large local Hilbert-space dimensions are achieved using the local nuclear-spin degree of freedom of the dopants, which range from $ \tfrac{1}{2}$ (P), over $\tfrac{7}{2}$ (Sb)~\cite{yu2024creation} to $\tfrac{9}{2}$ (Bi) depending on the choice of dopant atom. This platform has recently been used to simulate small instances of the extended Fermi-Hubbard model in (2+1)D~\cite{Salfi2016,le2017extended,Crane_phd,wang2022experimental} and the Su-Schrieffer-Heeger model~\cite{Le2020,Kiczynski2022}.

These results, in conjunction with similar demonstrations \cite{kiczynski2022engineering,hsueh2023hyperfine}, show that scanning tunneling microscopy (STM) lithography can be used to fabricate and probe arrays of dopants with sub-nanometer precision for applications in quantum simulation~\cite{Crane2018, stock2020}. Crucially, the nuclear and electron spins in these systems show extremely long coherence times~\cite{Watson2017} and, following theoretical proposals~\cite{Kane1998,PhysRevB.72.045350,OGorman2016,Veldhorst2017,Crane2019,Crane2021}, high-fidelity quantum-information-processing applications have been demonstrated experimentally~\cite{He2019,Ludwik2023,Thorvaldson24}. Dopants in silicon, therefore, exhibit the coherence properties required for high-fidelity quantum simulations.  Although current technology is limited to relatively modest lattice sizes on the order of ten sites, progress in fabrication techniques, particularly those demanding atomic precision, have been demonstrated~\cite{schofield2003atomically, stock2020, wyrick2019atom}. Moreover, advancements in external gates enable local electrical control of the donor charge states~\cite{He2019,Thorvaldson24}.

We show how to map a rotor Jackiw-Rebbi model to a donor platform (see Fig.~\ref{Fig:1}), and estimate the parameter regimes in which a faithful simulation can be achieved. We perform exact and approximate (using the Hartee-Fock-Bogoliubov method) numerical simulations to predict phenomena that could be observed in (1+1)D and (2+1)D. In this way, we map out the phase diagram of the model and identify the occurrence of dynamical mass generation, confinement-deconfinement transition, and the effect of thermal fluctuations on chiral-symmetry restoration. Finally, we discuss how these phenomena could be probed in donor-array simulators by local charge sensing and transport measurements. Our experimental and theoretical advances set the stage for future quantum simulations of lattice gauge theories in donors in silicon.

This paper is organized as follows. In Sec.~\ref{Sec:lattice}, we introduce the rotor Jackiw-Rebbi model, the donor platform along with its intrinsic Hamiltonian, and the reasoning for why this platform can be thought of as an analog quantum simulator of the rotor Jackiw-Rebbi model. In Sec.~\ref{Sec:1+1d}, we investigate the properties of the (1+1)D system and benchmark our results against the findings of Ref.~\cite{gonzalez2020rotor}. We further investigate the effect of Coulomb interactions on phases of the model.
Next, in Sec.~\ref{Sec:2+1d}, we delve into the phase diagram of the (2+1)D array and explore the phenomenon of dynamical mass generation in this extended model. Finally, in Sec.~\ref{Sec:Experiment}, we discuss the experimental feasibility of detecting the discussed phases in donor arrays by focusing on suitable observables.

\section{Donors in silicon as an analog quantum simulator of the rotor Jackiw-Rebbi model}\label{Sec:lattice}
\noindent
In this work, we lay out the reasoning for why donor arrays in silicon provide benefits for the simulation of (lattice) quantum field theories containing fermions and bosons. In particular, we focus on the physics of fermionic systems with a dynamical mass generated from interactions with bosonic degrees of freedom. As a first step, we focus on a proposal for simulating the rotor Jackiw-Rebbi model in (1+1)D and (2+1)D. A proposal for implementing the (1+1)D model using ultra-cold atoms in optical lattices was developed in Ref.~\cite{gonzalez2020rotor}, which, nonetheless, involved mapping the model to a fully bosonic quantum simulator. In contrary, the fermion-boson simulator of this work provides a more natural path to quantum simulation of the rotor Jackiw-Rebbi model in future implementations.

\subsection{Model Hamiltonian
\label{sec:model}}
Consider a massless Dirac-fermion field $\Psi$ and its adjoint $\overline{\Psi}=\Psi^\dagger \gamma^0$ in $d+1$ spacetime dimensions with $d=1,2$. The Dirac gamma matrices satisfy $\{\gamma^\mu,\gamma^\nu\}=2g^{\mu\nu}$ (under spacetime metric $g^{00}=1,g^{ii}=-1$ with $i=1,\cdots,d$). Consider also the vector fields $\bm{\phi}$ and $\boldsymbol{\ell}$, representing the orientation of a quantum rotor and its angular momentum, respectively. As will be explained in Sec.~\ref{sec:connection} and Appendix~\ref{app:CL}, in an equivalent lattice formulation, $\bm{\phi}$ captures the N\'eel alternation of antiferromagnetic ordering between spins at neighboring lattice sites while the presence of $\boldsymbol{\ell}$ also allows for slow spin fluctuations with respect to uniform ordering. The continuum Hamiltonian of a corresponding lattice model we aim to study is:
\begin{align}
\label{eq:H-RJR-d+1}
H^{\rm{(RJR)}}=
\int d^dx \,
\bigg\{& \bigg[-ic \overline{\Psi}(\bm{x}) \gamma^i \partial_i \Psi(\bm{x})\nonumber\\
&-\overline{\Psi}(\bm{x}) \big[\boldsymbol{g}\cdot \bm{\phi}(\bm{x})+ \gamma^0 \boldsymbol{g}\cdot\boldsymbol{\ell}(\bm{x})\big]\Psi(\bm{x})\nonumber\\
&-\bm{h} \cdot \bm{\ell}(\bm{x}) \, \biggr]\bigg\}.
\end{align}
The first term in Eq.~(\ref{eq:H-RJR-d+1}) is the free-fermion Hamiltonian, the second term describes the interactions of fermions with bosonic quantum rotors via the coupling vector $\bm{g}$, and the last term is responsible for rotor's precession in a uniform magnetic field $\bm{h}$. In (1+1)D, the Hamiltonian is invariant under a discrete chiral transformation, $\Psi \to \gamma^5 \Psi$, with $\gamma^5=i\gamma^0 \gamma^1=i\sigma^z$, if $\phi = 0$. Such a symmetry gets broken spontaneously if $\langle \phi \rangle \neq 0$, hence generating a dynamical fermion mass in the model. In (2+1)D, no chirality can be defined but the Hamiltonian with $\phi = 0$ is invariant under another discrete transformation, i.e., spatial reflection, that only changes the sign of one of the two components of the position vector, e.g., $x_1 \to -x_1$, transforming the fermion field as $\Psi \to \gamma^1 \Psi$. Again, if $\langle \phi \rangle \neq 0$, such a symmetry gets broken spontaneously, generating a dynamical fermion mass.

The model described by the Hamiltonian in Eq.~(\ref{eq:H-RJR-d+1}) can be identified as a variant of the Jackiw-Rebbi model, as introduced in Ref.~\cite{gonzalez2020rotor}. In the original Jackiw-Rebbi model~\cite{jackiw1976solitons}, fermions interact with a self-interacting scalar field. In the model considered here, called rotor Jackiw-Rebbi model, fermions interact with rotor fields, and these rotors exhibit no self interactions. This distinction is significant: in the Jackiw-Rebbi model, the contribution from the scalar field to the generation of dynamical mass through symmetry breaking can be explained classically, but this is not the case in the rotor Jackiw-Rebbi model. This model, in fact, bears a stronger resemblance to the Gross-Neveu models, although without an $O(\mathbb{N})$ symmetry~\cite{gross1974dynamical}. 

Due to intrinsically quantum and nonperturbative spontaneous symmetry breaking and dynamical mass generation in the rotor Jackiw-Rebbi model, one needs nonperturbative simulation tools to study model's phases. In this model, spontaneous symmetry breaking turns out to occur within a specific range of external magnetic fields. As is shown in Ref.~\cite{gonzalez2020rotor} for the (1+1)D model, the range is notably dependent on a function which cannot be approximated by a Taylor expansion for small but nonzero rotor-fermions coupling. A similar behavior is expected in (2+1)D, as is studied in this work. In the subsequent sections, we will explore in detail how a dopant analog quantum simulator can effectively implement a lattice-discretized version of the rotor Jackiw-Rebbi model in both (1+1)D and (2+1)D.

\subsection{Simulator Hamiltonian
\label{sec:simulator}}
To investigate how the proposed dopant array, as illustrated in Fig.~\ref{Fig:array_schematic}, can be used to probe the physics of dynamical mass generation in the rotor Jackiw-Rebbi model, we begin by introducing the platform and its intrinsic Hamiltonian.

To prepare a donor array, first, standard semiconductor processing techniques are used to create a clean surface. The dopant atoms are then incorporated into the surface using scanning-tunneling-microscope hydrogen lithography techniques~\cite{usman2016spatial,wang2020atomic}. Following the incorporation of the dopants, silicon is grown over the array to encapsulate the donors and hold them in place, as well as allow further lithography of metallic top gates. This protected and contollable lattice of dopants then creates a new band structure within the band gap of the underlying silicon crystal. Electrons moving in this dopant lattice can, therefore, be seen as an analog to cold atoms hopping in an optical lattice, with the silicon lattice only participating as a dielectric, but otherwise being inert. This analogy has opened up the demonstration of this platform as a novel native fermionic quantum simulator with an early application to simulate the Fermi-Hubbard model~\footnote{The effective donor Hubbard model in silicon crystals has parameters significantly smaller than those in the crystal itself. This means modest physical temperatures correspond to extremely low temperatures in Hubbard parameter units. For instance, the on-site Coulomb interaction is less than $400$ K. Additionally, the capacity for transport measurements and strong, adjustable long-range Coulomb interactions make this platform a viable alternative to studying the Hubbard model, compared to cold atoms in optical lattices.} and beyond~\cite{Salfi2016,le2017extended, wang2022experimental,kiczynski2022engineering}. Furthermore, electrons interact with themselves and with the dopant nuclei via Coulomb interactions, and both electron and nuclear spin respond to external magnetic fields. Importantly, the electron spin couples to the nuclear spin via a hyperfine interaction.

Let us discuss these interactions in more detail. The intrinsic Hamiltonian of the dopant analog quantum simulator can be decomposed to
\begin{equation}\label{eq:1}
H^{\rm{(lattice)}}=H_\mu+H_\text{k}+H_{\text{ff}}+H_{\text{sf}}+H_{\text{s}}.
\end{equation}

The chemical-potential term, denoted as \( H_\mu \), represents the onsite binding energy, \( E_b \), of a single electron to a dopant in the absence of tunneling. By carefully adjusting the energy level of the system to be near the Fermi level, one can control the total electron occupancy in the system. The other contribution to the onsite energy term comes from the local Coulomb potential between the electron and the nuclei of the dopants within the array. Additionally, there is another contribution to the overall electric potential, resulting from the gates and other charges surrounding the system, denoted by \( \epsilon \)~\cite{le2017extended}. Overall, \( H_\mu \) can be written as
\begin{equation}\label{Eq:2}
H_\mu=\sum_{i}\sum_{\sigma}\bigg[\big(-E_{b,i}+\epsilon_{i}\big)c_{i,\sigma}^\dagger c_{i,\sigma}+\sum_{j \ne i}V_{0,ij}c_{i, \sigma}^\dag c_{i,\sigma}\bigg].
\end{equation}
Here, $ c_{i,\sigma}^\dagger$ and $c_{i, \sigma}$ represent the electron's creation and annihilation operators, respectively. $\sigma = \, \uparrow,\downarrow$ is the spin index of the electron, and $i$ denotes the site index within the $d$-dimensional array with given geometry. For example, this index ranges from $1$ to $N=N_s^d$ on a square lattice with side length $L=aN_s$ and lattice spacing, or constant, $a$. $V_{0,ij}$ represents the strength of electric potential at site $i$, created by the nucleus at site $j$. We approximate the interaction potential by $V_{0,ij}=V_0e^{-\lambda |\bm{r}_i-\bm{r}_j|}/|\bm{r}_i-\bm{r}_j|$, where the constant term $V_0=1/(4\pi \epsilon_0 \epsilon_{\text{si}})$ reflects the permittivity of the electric field in silicon, $\epsilon_{\text{si}}$, and $\lambda$ determines how the electric field is reduced due to screening ~\cite{le2017extended,wang2022experimental,kim2020control}. $\bm{r}_i$ denotes the position vector of site $i$.

The kinetic term $H_\text{k}$ describes the hopping of electrons between neighboring sites in the array. In this study, we assume that the hopping strength can be engineered to be uniform across the system, which demands an atomically precise placement of the donors. In general, the hopping strength decays exponentially with distance, so we restrict this study to nearest-neighbor hopping of strength $t$:
\begin{equation}
    H_{\text{k}}=-t\sum_{\langle i,j\rangle}\sum_{\sigma} c_{i,\sigma}^\dagger c_{j,\sigma}+\text{h.c.}
\end{equation}

$H_{\text{ff}}$ represents the fermion-fermion interaction in the system. This interaction can be modeled as the sum of two components: the Coulomb interaction between fermions across the array and the onsite interaction between two electrons with opposite spins. The strength of the latter interaction is denoted by $U_i$:
\begin{equation}\label{Eq:H_ff}
H_{\text{ff}}=\sum_i \bigg[\sum_{j\neq i}\sum_\sigma V_{0,ij}
n_{i,\sigma}n_{j,\sigma}+U_i n_{i,\uparrow}n_{i,\downarrow}\bigg], 
\end{equation}
where $n_{i,\sigma}\coloneqq c_{i,\sigma}^\dagger c_{i,\sigma}$ is the number operator for a fermion with spin $\sigma$ at lattice site $i$. 

$H_{\text{sf}}$ represents the most crucial term of relevance to the model we aim to study. It involves the interaction of the dopant nuclear spin $\boldsymbol{I}$ with the electrons,
\begin{equation}
    H_{\text{sf}}=
    g\sum_{i} s_i^z I_i^z.
\end{equation}
Here, the parameter $g$ represents the hyperfine coupling, and $s_i^z \coloneqq  c_{i,\uparrow}^\dagger c_{i,\uparrow}-c_{i,\downarrow}^\dagger c_{i,\downarrow}$ is the electron-spin operator along the $z$ direction. The electrons will be assumed to be spin-polarized in this direction by an external magnetic field, $\boldsymbol{B}$. 

We allow for a transverse magnetic field, $h_x$, arising from time-dependent Nuclear Magnetic Resonance (NMR) control. This leads to an effective Hamiltonian in the rotating frame of the nuclear spins:
\begin{equation}
\label{eq:Hs-dopant}
    H_{\text{s}}=-\sum_{i} \boldsymbol{h} \cdot \bm{I}_{i} +\sum_{i} h_e s_i^z.
\end{equation}
Here, the first term is the rotating-frame Hamiltonian for the nuclear spins in the presence of the NMR drive with the magnetic field in the form of $\bm{h}= h_z \hat{\bm{z}} +h_x \hat{\bm{x}}$. We will work in the limit of a large drive detuning, where $|h_z| \gg |h_x|$. Given this condition, we omit the effect of the transverse magnetic field, $h_x$, on the electrons. We assume that in this regime, the spin polarization of electrons due to the transverse magnetic field remains constant and is negligible. Finally, $h_e$ is the Zeeman splitting of the electron spin. 

In the above Hamiltonian, we work in a configuration where the electrons' spins are polarized in the $\ket{\downarrow}$ direction (see Sec.~\ref{Sec:Experiment} for how to create such a polarization). 
In this case, the spin drops out of the Hamiltonian terms and one gets
\begin{align}\label{Eq:lattice_model}
\begin{split}
    H^{\rm{(lattice)}}=&-t\sum_{\langle i , j\rangle} 
    \left(c_{i}^{\dagger} c_{j}+\text{h.c.}\right)-\sum_{i} \mu_{i} 
    c_i^\dagger c_i\\
    &-\sum_{i}\left(g c_{i}^{\dagger} c_{i} I_i^z+\boldsymbol{h} \cdot \boldsymbol{I}_{i}\right)\\
    &+\sum_{i}\sum_{j\neq i} V_{0,ij}c_i^\dagger c_i c_j^\dagger c_j.
\end{split}
\end{align}
Here, $c_i$ ($c_i^\dagger$) denotes annihilation (creation) operator for the down-spin electron at site $i$ and $\mu_i = E_{b,i}-\epsilon_{i}-\sum_{j \ne i}V_{0,ij}$. Note that since the total number of electrons in each spin state is conserved, the last term in Eq.~(\ref{eq:Hs-dopant}) contributes trivially to the dynamics and is dropped. In the subsequent sections, we explore the connection between this lattice Hamiltonian and the rotor Jackiw-Rebbi model.

\subsection{The connection between the Rotor Jackiw-Rebbi model and the lattice model
\label{sec:connection}}
Standard techniques can be employed to map the low-energy limit of a lattice theory of fermions and spins to a continuum quantum field theory, see Ref.~\cite{gonzalez2020rotor} and Appendix~\ref{app:CL}.

Let us first consider the free-fermion Hamiltonian. Here, we assume a uniform electron chemical potential throughout the lattice such that $\mu_i \coloneqq  \mu$ is constant. We further assume a fermionic state with half filling, i.e., $\tfrac{1}{N}\sum_i \langle c_i^\dagger c_i\rangle = \tfrac{1}{2}$. There is a quantum field theory describing low-energy excitations around the Fermi points (surface). In (1+1)D, this means linearizing the dispersion relation  $\epsilon(k)=-2t\cos(ka)$ near two (free-fermion) Fermi points at $k_F=\pm \tfrac{\pi}{2a}$, i.e., $\epsilon(k)\approx (|k|-k_F)v_F$, with the Fermi velocity defined as $v_F\coloneqq 2ta$. By this approximation, the fermionic operators $c_i$ in (1+1)D are reduced to left- and right-moving slowly varying fermion fields, $\psi_+(x)$ and $\psi_-(x)$. These fields map to the staggered lattice fermions of Kogut and Susskind~\cite{kogut1975hamiltonian} (see Ref.~\cite{gonzalez2020rotor}), which is a lattice discretization of the continuum Dirac fermions in Eq.~(\ref{eq:H-RJR-d+1}), upon the identification $\Psi = (\psi_+~\psi_-)^T$ and $c=v_f$.

On a (2+1)D square lattice, the (free-fermion) Fermi surface is the boundary of the diamond shape $-\tfrac{\pi}{\sqrt{2}a}<k_\pm\leq \tfrac{\pi}{\sqrt{2}a}$ within the Brillouin zone, where $k_\pm = \tfrac{1}{\sqrt{2}}(k_x \pm k_y)$. Linearizing around this boundary does not give rise to the Dirac fermions of the continuum as is known. Nonetheless, as shown in Appendix~\ref{app:CL} and Sec.~\ref{Sec:2+1d}, due to the Fermi Surface admitting a nesting vector at momentum $(\pi/a,\pi/a)$, a N\'eel-order phase arises in this model too, generating dynamical mass for the fermions. The two-component $\Psi$ field can be understood as pairs of lattice fermions on adjacent sites along the $x$ direction, hence breaking the reflection symmetry~\footnote{Or in this case, translational symmetry along the $x$ direction.} discussed in Sec.~\ref{sec:model} in the N\'eel phase, see Appendix~\ref{app:CL}. 
Alternatively, to recover the continuum Dirac theory in Eq.~(\ref{eq:H-RJR-d+1}), fermions can be placed on a honeycomb lattice. On such a lattice, upon the linearization of the dispersion function, one arrives at $\varepsilon(\bm{k})=- v_F\bm{\kappa} \cdot \bm{\gamma}$, where $\bm{\kappa}$ is a small momentum around the Fermi points $\bm{K}$ and $-\bm{K}$ with $\bm{K}=\tfrac{2\pi}{3a}(1, \tfrac{1}{\sqrt{3}})$. This recovers the Dirac free-fermion theory upon the identification $c=v_F$ and
\begin{align}
    \label{eq:defpsi1}
    \Psi&=\begin{pmatrix}
    c_{\pm \bm{K}+\bm{\kappa},A}\\
    -c_{\pm \bm{K}+\bm{\kappa},B}\end{pmatrix}
\end{align}
around the Fermi points $\pm\bm{K}$. Here, $c_{\bm{k},A}$ and $c_{\bm{k},A}$ are the Fourier transform of the $c_i$ operators defined on sublattices $A$ and $B$ of the honeycomb lattice, respectively (see Appendix~\ref{app:CL} for details).

To achieve a low-energy quantum field theory of nuclear spins on the lattice, i.e., a long-wavelength large-spin limit, a continuum field of unit length, $\bm{\phi}$, can be introduced such that it varies slowly on the scale of a lattice spacing. To allow for a quantum disordered phase with no long-range spin order, it is imperative to also include a component of the spins which is perpendicular to the local orientation of the N\'eel order, which is denoted by the continuum field $\bm{\ell}$, as discussed in Appendix~\ref{app:CL}. Formally, this can be described as
\begin{equation}
    \bm{I}(\bm{r}_i) \simeq 
    S\left[\delta_i \bm{\phi}(\bm{r}_i)\sqrt{1-a^{2d}\bm{\ell}^2(\bm{r}_i)}+a^d \bm{\ell}(\bm{r}_i)\right]
    \label{Eq:nspindecomp}
\end{equation}
such that $\bm{\phi}\cdot \bm{\phi}=1$, $\bm{\phi}\cdot \bm{\ell}=0$, and $a^{2d}\bm{\ell}^2 \ll 1$. Here, $\delta_i$ is either $1$ or $-1$ depending on which sublattice, $A$ or $B$, $\bm{r}_i$ belongs to, see Ref.~\cite{sachdev1993quantum} for more details.

As shown in Appendix~\ref{app:CL}, the low-energy excitations of fermions near the Fermi surface are mediated by the rotors in either the N\'eel-order phase or the disordered phase. Hence, the hyperfine coupling of nuclear spins to electrons maps to the fermion-rotor interactions in the rotor Jackiw-Rebbi model. Additionally, the nuclear-spin interactions in presence of the NMR drive are naturally reduced to the last term in $H^{\rm{(RJR)}}$ in Eq.~\eqref{eq:H-RJR-d+1}. Finally, while electron-electron Coulomb interactions do not have an equivalent low-energy description in the rotor Jackiw-Rebbi model, as our numerics show, these interactions do not affect qualitative features of the phases of the model and the dynamical mass generation when experimental parameters are concerned.

With the connection between the (discretized) rotor Jackiw-Rebbi model and the lattice model accessible in the quantum simulator established, we are ready to delve into the study of the properties and different phases exhibited by the lattice model.

\section{Phenomenological study of the model in (1+1)D
}\label{Sec:1+1d}
\begin{figure}[t!]
\includegraphics[width=0.5\textwidth]{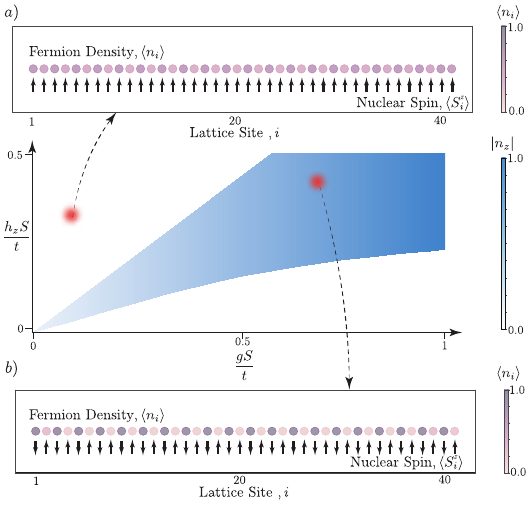}
\centering
\caption{\label{Fig:1d_results}
\textbf{Phase diagram of a (1+1)D system.} The phase diagram obtained from Hartree-Fock simulations, illustrates two distinct phases, trivial (a) and nontrivial (b), influenced by external magnetic fields $h_z$ at fixed $h_x$. The nontrivial N\'eel phase, indicated in blue, emerges within a specific range of $h_z$ when $h_x \ll h_z$, while the trivial phase, where all spins are polarized in a single direction, is represented by a white background. The circles represent the density of fermions at each site, where a darker color indicates higher density and the arrows show the orientation of the dopant's spin. The blue color represents the parameter $|n_z|$, where $n_z$ is defined in Eq.~\eqref{Eq:nz-def}, which is ideally one for the N\'eel-order phase and zero for the trivial phase. The simulation is done on a one-dimensional array with $N=43$ sites in a configuration with one fermion above the half-filling state. The arrows represent the spin, $\langle S^z_i \rangle$, of the dopants, while the circles denote the fermion density at each site, indicated as $\langle n_i \rangle$. The parameter $h_xS/t$ is set to $0.01$, where $h_x$ denotes the external magnetic field in the $x$ direction and $t$ denotes the hopping parameter. For this plot, we have set $t=7.5~\text{meV}$ and $V_0/(at)=1.1$} 
\end{figure}
\noindent
To investigate the quantum many-body system described by the simulator Hamiltonian in Eq.~\eqref{Eq:lattice_model}, one might consider employing exact diagonalization (ED) techniques. However, the dimensionality of this Hamiltonian grows exponentially with the system size, making the application of ED with available classical computational resources feasible only up to \( O(10^1) \). To produce more accurate high-\( L/a \) results, studies on system with larger sizes are required. While well-known efficient methods like density-matrix renormalization group (DMRG)~\cite{white1992density} and tensor networks~\cite{hauschild2018efficient} are suitable for the (1+1)D cases, extending these methods to (2+1)D has proven challenging. To address this challenge, albeit with a modest sacrifice in accuracy, we have employed the mean-field approach of Hartree-Fock (HF)~\cite{giuliani2005quantum} and finite-temperature Hartree-Fock-Bogoliubov (FTHFB)~\cite{goodman1981finite}. These enable the exploration of larger systems with a computational complexity proportional to the system size \( O(N) \). Comparing the results obtained from the HF approximation with those from ED for systems up to \(O(10)\) lattice sites, we have nonetheless demonstrated that the mean-field approach is a reliable method of approximation, even when considering small values of dopant spin, \(S\). As the system size increases, i.e., in the limit of large \(L/a\) and large \(S\), a previous study~\cite{gonzalez2020rotor} shows that the mean-field predictions can converge to the exact methods smoothly.

\subsection{Dynamical mass generation
}\label{Sec:dyn-mass-1+1d}
To study the distinct phases of the lattice model in Eq.~\eqref{Eq:lattice_model}, we can explore the phase space in terms of a set of macroscopic parameters, including hyperfine coupling and the external longitudinal and transverse magnetic fields, which are denoted respectively as \( (g, h_z, h_x) \). We will also investigate the impact of these parameters on the transition between possible phases for given fixed values of dopant spin, tunneling coupling, the temperature of the reservoir surrounding the system, Coulomb-force coupling, chemical potential, and lattice constant, denoted respectively as \( (S, t, T, V_0, \mu, a) \). The list of parameters has been provided in Appendix.~\ref{paramters}. The boundary condition is another factor that needs to be set in the simulation. We have investigated the effects of open and periodic boundary conditions and found that the periodic one can better mitigate the (undesired) effects of Coulomb interactions. Thus, in this paper, we present results concerning the periodic case. 

Our numerical simulations reveal that modulating the macroscopic parameters gives rise to two distinct phases within the system, characterized by the phenomenon of dynamical mass generation. Building upon the method described in Ref.~\cite{gonzalez2020rotor}, we utilize the modified N\'eel-order parameter, denoted as $n_z$, as the principal observable to investigate spontaneous symmetry breaking in the phase diagram and to distinguish between different phases. The parameter $n_z$ is defined as
\begin{align}
\label{Eq:nz-def}
n_z \coloneqq \frac{1}{N}\sum_{i=1}^N\frac{1}{S}\left(\langle S_1^zS_{2i}^z\rangle-\langle S_1^zS_{2i-1}^z\rangle\right),
\end{align}
where $i$ denotes the index of the lattice site, $N$ is the total number of lattice sites, and $S_i$ represents the spin of the dopant at location $i$ in the array. Our numerical simulations, conducted using the mean-field HF method, demonstrate the existence of two distinct phases for the (1+1)D lattice model. This phase diagram is presented in Fig.~\ref{Fig:1d_results} (see Appendix \ref{app:supp_plots} for more details). The phase diagram is derived at zero temperature and one electron above the half-filling state, for a wide range of $gS/t$ and  $h_zS/t$, and a fixed nonzero transverse magnetic field $h_x$ and given the Coulomb interaction with strength $V_0$.

The first phase, termed the `trivial phase', is characterized by all spins being polarized and aligned with the external magnetic field. This alignment results in a zero value for \( n_z \). In this configuration, the fermion density across the system is found to be distributed around half-filling for all lattice sites, indicating the absence of dynamical mass terms in Eq.~\eqref{eq:H-RJR-d+1}. In this phase, the fermions are massless, thereby exhibiting chiral symmetry. This state of the system is analogous to the Luttinger-Liquid (LL) phase for fermions in (1+1)D. In contrast, the `nontrivial phase' emerges within a specific range of macroscopic parameters, where \( n_z \) departs from zero. The corresponding fermion density across the system indicates the presence of dynamical mass generation. As discussed in the preceding section, chiral symmetry is spontaneously broken in this phase, and the system settles into one of the ground states.
\begin{figure}
    \centering\includegraphics[width=0.5\textwidth]{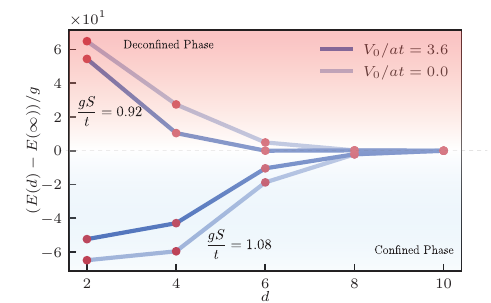}
    \caption{\textbf{Confinement/deconfinement phase transition.} Plots show the normalized system energy, $E$, as a function of the pinning-potential range, $d$, minus the normalized energy in the limit $d \gg 1$. Adjusting $gS/t$ in a certain range with fixed $h_z$ creates quark bags in the N\'eel phase, increasing energy as $d$ increases (quark confinement). Lowering $gS/t$ below a threshold results in a deconfined phase with reduced energy. Coulomb interaction present in the silicon-based experimental platform does not alter this phenomenon.  The numerical simulation is based on the Hartree-Fock method at half-filling with $h_xS/t=0.01$, $h_zS/t=0.4$, $a=4.7~\text{nm}$, and $t=7.5~\text{meV}$}\label{Fig:Ed}
\end{figure}
%

\subsection{Confinement/deconfinement  of fractionally charged quasi-particles}
To investigate the phenomenon of confinement and deconfinement of fractionally charged fermions 
coupled to the N\'eel field, it is essential to introduce a pinning potential as an impurity to break the translational symmetry. By adopting the method detailed in Ref.~\cite{gonzalez2020rotor}, we add a pinning potential, \(H_\epsilon=-\sum_{i=1}^{N}\varepsilon_i(d) S_i^z\), which can lead to fermions localizing at the lattice positions \(i_0\) and \(i_0+d\) for a certain range of \(h_zS/t\) and \(gS/t\). Here,
\begin{equation}
\varepsilon_i(d) =
\begin{cases} 
\varepsilon (-1)^i & 1 \leq i < i_0 \\
\varepsilon (-1)^{i+1} & i_0 \leq i < i_0 + d \\
\varepsilon (-1)^i & i_0 + d \leq i \leq N
\end{cases}
\end{equation}
for small pining potential strength $\varepsilon/t=0.05$ compared to the tunneling coupling. Since the total extra fermionic density is equal to one electron, one can refer to this configuration as fractional charge since at each end, there is half of an electron localized.

To examine the system's dynamics, we calculate the static potential between two fractional charges, \(V(d)=E(d)-E_\infty\), where \(E_\infty\) denotes the ground-state energy when \(d\gg 1\). This analysis unveils two distinct, confined and deconfined, phases. The confined phase features an ascending static potential with respect to \(d\), indicating the binding of fractional charges. In this model, the integer-charged fermion field couples to the rotors field, allowing, under certain conditions, the N\'eel field's configuration to host localized fractional charge, mirroring quark bags in the phenomenological models of QCD, where the quarks are confined within a finite region of space known as the `bag'~\cite{chodos1974baryon,johnson1975bag} and make up pions, nucleons, and other hadrons. Conversely, the deconfined phase shows a decreasing static potential, signifying no binding between fractional charges. Figure~\ref{Fig:Ed} illustrates these phases, emphasizing that they can occur even amidst Coulomb repulsion in a silicon platform, highlighting the unique interplay between fractional charges and the N\'eel-field configuration. In this plot, we choose two coupling strengths: one assuming the Coulomb interaction does not exist, $V_0/(at)=0$, and the second one with the highest value of $V_0/(at)=3.6$ for silicon with fixed $a=4.7~\text{nm}$ and $t=7.5~\text{meV}$, which are chosen close to experimentally favorable numbers.

\section{Phenomenological study of the model in (2+1)D}\label{Sec:2+1d}
\noindent
In the subsequent section, we examine evidence for the occurrence of dynamical mass generation and spontaneous symmetry breaking within the (2+1)D model. As mentioned before, the continuum physics of square and honeycomb lattice models are distinct, and only the latter recovers Dirac fermions of the Jackiw-Rebbi model. Nonetheless, we study both cases in the following.

\subsection{
Dynamical mass generation on a square lattice
}\label{Sec:dyn-mass-2+1d}
\begin{figure}[t!]
\includegraphics[width=0.5\textwidth]{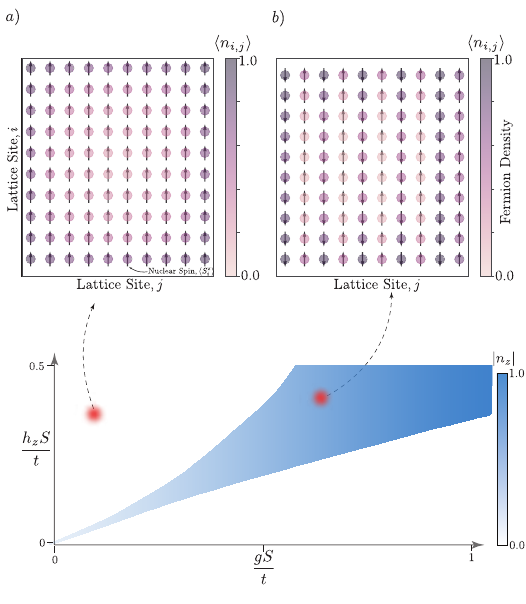}
\centering
\caption{\label{Fig:2d_skectch} \textbf{Phase diagram of a (2+1)D system.} Similar to the (1+1)D scenario in Fig.~\ref{Fig:1d_results}, the spin configuration and fermion distributions in a two-dimensional square array, $N=10 \times 10$, are depicted. The arrows represent the spin, $\langle S_z \rangle$, of the dopants, while the circles denote the fermion density at each site, indicated as $\langle n_{i,j} \rangle$. Note that we have now introduced a two-dimensional index $(i,j)$ compared to the linear index $i$ introduced in Sec.~\ref{sec:simulator}. The phase diagram of the system, in terms of macroscopic parameters, comprises two phases: the trivial phase (a) and the N\'eel-order phase (b). The blue color represents the parameter $|n_z|$ defined in Eq.~\eqref{Eq:nz-def}, which is ideally one for the N\'eel-order phase and zero for the trivial phase.
For this plot, we have set $t=7.5~\text{meV}$, $h_xS/t=0.01$, and $V_0/(at)=1.1$
}
\end{figure}

We apply both ED for small systems and the HF method for larger systems to investigate phases of the lattice model on a square lattice for a range of macroscopic parameters. Our simulations demonstrate that a phase diagram akin to the one observed in the (1+1)D lattice can also be observed in (2+1)D. Figure~\ref{Fig:2d_skectch} illustrates
the phase diagram for a $N=10 \times 10$ array obtained using the mean-field HF method. 

To extend our study, we further consider the system at a finite temperature, $T\coloneqq 1/\beta$, hence consider the thermal expectation value of the total number of fermions, denoted as 
$\sum_{i}\langle  c_i^\dagger c_i\rangle_\beta $, where expectation value is with respect to the Boltzmann factor $e^{-\beta H^{\text{(lattice)}}}$.
This expectation value, when considering  the  grand canonical ensemble at a fixed temperature, becomes 
a function of the chemical potential, represented as $\mu$.

As shown in Appendix \ref{appendix_FTHFB}, achieving a half-filling state requires independent tuning of the chemical potential for a given temperature. 
Within our HF calculation, we have also checked if there is nonzero pairing in the (Bardeen–Cooper–Schrieffer) BCS channel. To start with, we analyzed the quantity $\tilde{K} \coloneqq \frac{1}{N^2}\sum_{i,j} K_{ij} K_{ij}^*$ where $K_{ij} \coloneqq \langle c_i c_j \rangle_\beta$. The results, as shown in Fig.~\ref{Fig:tr_rho,tr_K} in Appendix~\ref{appendix_FTHFB}, demonstrate that the averaged pairing correlator $\langle c_i c_j \rangle_\beta$, remains zero throughout the entire range of chemical potentials and temperatures considered. 
However, the approximate HF method employed effectively only looks at Coulomb repulsion as a cause for pairing. The possible role of interactions between nuclear-spins and electrons in pairing is not addressed with this approximate method. Other studies ~\cite{he2023superconductivity,crepel2021new,crepel2022unconventional} with a Hamiltonian that shares similar terms as our system indicate that pairing can occur at a certain range of Coulomb coupling and hyperfine coupling $V_0/g \ll 1$. This range is the opposite of the physical condition in the analog quantum simulator of this work, i.e., $V_0/g \gg 1$, and is hence not analyzed further.

\begin{figure}[t!]
\includegraphics[width=0.5\textwidth]{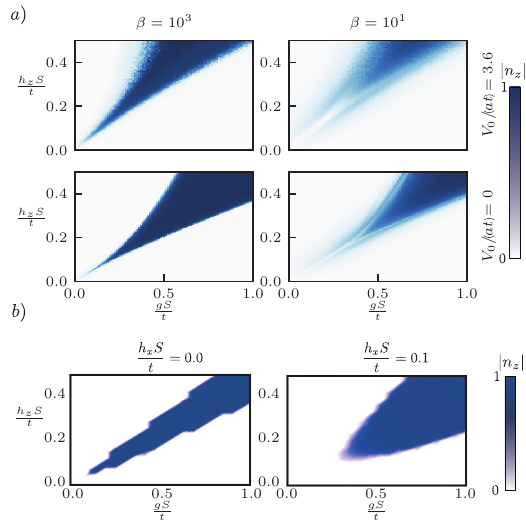}
\centering
\caption{ \label{Fig:beta_V0_effect}\textbf{The impact of temperature, Coulomb potential, and transverse magnetic field on the phase diagram of the (2+1)D system.}
a) Comparing the subplots in the left and right columns illustrates the influence of increasing temperature on the phase diagram. Additionally, comparing the top and bottom rows sheds light on the effect of Coulomb's repulsion force on the phase diagram. The results are based on the Finite-Temperature Hartree-Fock method for an array of $N=10\times 10$, and the chemical potential is tuned such that the system is around half-filling. The blue color represents the parameter $|n_z|$ defined in Eq.~\eqref{Eq:nz-def}.
b) The different phases of the square array are plotted for two different transverse fields, $h_x$. The simulations for b) are performed using ED for the total number of spin $N=2\times 2$ at half-filling. The Coulomb potential is set to $V_0/(at)=3.6$. The color represents the absolute value of the order parameter, $|n_z|$. In both a) and b), $t=7.5~\text{meV}$, and in part a), $h_xS/t=0.01$.}  
\end{figure}

Our investigation, according to Fig.~\ref{Fig:beta_V0_effect}, further reveals that as the temperature deviates from absolute zero, nontrivial phases gradually diminish, ultimately driving the quantum state towards a thermal state, as predicted. Under such circumstances, discrete reflection symmetry in (2+1)D is restored as the temperature rises, causing the fermionic field to revert to a zero-mass state. A similar effect can also be partially induced through the escalation of Coulomb interactions. Nonetheless, for the coupling strength \(V_0\) from zero to the highest value of \(123 \text{ nm}\cdot \text{meV}\) for silicon, the nontrivial N\'eel-order phase persists, even when considering nonzero temperatures. Finally, the transverse component of the external magnetic field, \(h_x\), introduces fluctuations to the system. As illustrated in Fig.~\ref{Fig:beta_V0_effect}, increasing \(h_x\) has an effect qualitatively analogous to raising the temperature on the restoration of discrete reflection symmetry.

\subsection{Dynamical mass generation on a honeycomb lattice
}
\begin{figure*}[t!]
\includegraphics[width=\textwidth]{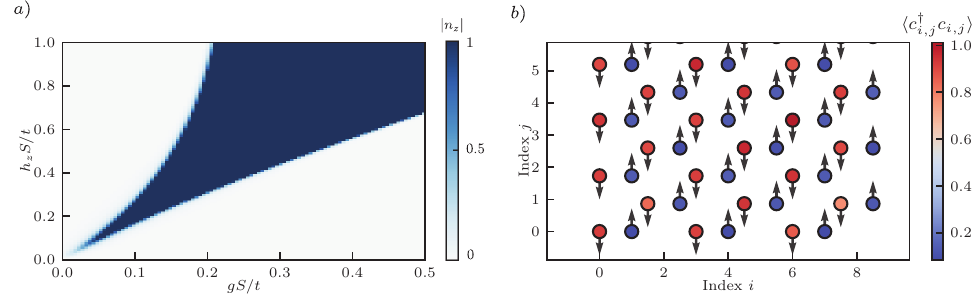}
\centering
\caption{\label{fig:hex_phase_diagram}\textbf{Phase diagram of a (2+1)D honeycomb array for $N=72$ sites.} a) Similar to the square lattice shown in Fig.~\ref{Fig:2d_skectch}, the phase diagram in terms of macroscopic parameters comprises two phases: the N\'eel-order phase and the trivial phase, with the presence of Coulomb interaction. The blue color represents the parameter $|n_z|$ defined in Eq.~\eqref{Eq:nz-def}, which ideally is one for the N\'eel-order phase and zero for the trivial phase. 
b) The dopant-spin configuration and fermion-density distributions ($\langle c_{i,j}^\dagger c_{i,j} \rangle$) in the honeycomb lattice are shown for a region where the system exhibits the N\'eel-order phase. The numerical simulations are based on the finite-temperature Hartree-Fock method with inverse temperature $\beta=10^3$. The tunneling coupling and horizontal external magnetic field are set similar to those in the square lattice, with $t=7.5~\text{meV}$, $h_xS/t=0.01$, and $V_0/(at)=1.1$. The chemical potential is tuned to ensure that the system is around half-filling in the N\'eel-order phase.
}
\end{figure*}
In this section, we aim to broaden our study to include another structure in the $(2+1)$D case, specifically focusing on the honeycomb lattice. On one hand, the continuum model of fermions on a honeycomb lattice is more tractable---its tight-binding model admits Dirac fermions like graphene. 
On the other hand, this structure is of significant interest considering previous research that explored the emergence of superconductivity in the behavior of the system when the ratio of $V_0/g$ couplings falls within a certain regime~\cite{he2023superconductivity,crepel2021new,crepel2022unconventional}. In these lattice structures, phenomena similar to the confinement-deconfinement transitions observed in (1+1)D may be linked to superconductivity behavior and pairing mechanisms. Recent work in Ref.~\cite{PhysRevResearch.5.L012009} on a Hamiltonian which incorporates similar interactions has shown signs of 
unconventional superconductivity, exhibiting staggered d-wave and f-wave pairings, along with topological superconductivity. This phenomenon can be compared to the confinement phenomenon observed in the (1+1)D case. 

We apply our numerical procedure to the honeycomb lattice. Employing FTHFB simulations, while incorporating the presence of Coulomb interactions, we have examined the phase diagram for a wide range of $(gS/t, hS/t)$ parameters. We observe a pattern similar to the N\'eel order within a certain range of an external magnetic field as shown in Fig.~\ref{fig:hex_phase_diagram}.

Despite the observation of the two phases in this model, realizing the parameter regime relevant to the emergence of superconductivity~\cite{he2023superconductivity,crepel2021new,crepel2022unconventional} is very challenging because the hyperfine coupling that sets the strength of the sub-lattice potential needs to be comparable to the Coulomb-interaction strength. Instead, a more promising route to explore emergent superconductivity in this platform is to introduce a static sub-lattice potential using external gates or by controlling the number or type of dopant on each site. We leave the investigation of such extensions of our model as an interesting avenue for future work.

\section{Experimental Probing }\label{sec:experimental_probing}
\noindent
In the following section, we investigate the experimental viability of observing the phase transition described in earlier sections in present doped semiconductor quantum simulators. It is crucial to establish this viability, as the desired phase transitions occur within certain ranges of model parameters that should be matched to the ranges accessible by current laboratory capabilities. Additionally, we discuss methods for preparing the initial state and detecting the system's phase, as well as other relevant observables within our proposed framework for two-dimensional dopant arrays. We remark that all the necessary ingredients for our proposal have been previously demonstrated in experiments with phosphorus donors in silicon~\cite{pla2013high,He2019,Savytskyy23,Thorvaldson24}.

\subsection{State preparation}

The initial phase of an experimental investigation involves setting the system to a preferred initial state. In the magnetic-field range characterized by \( B \), with electron and nuclear gyromagnetic ratios \(\gamma_e\) and \(\gamma_n\), such that \( \gamma_e B \gg g > 2 \gamma_n B \), the eigenstates of the fermion--nuclear-spin system at each site, in the absence of hopping and Coulomb interactions, can be segmented into four distinct states.
These are ordered from the lowest energy level to the highest as \( \ket{\downarrow, \Uparrow } \), \( \ket{\downarrow, \Downarrow } \), \( \ket{\uparrow, \Downarrow } \), and \( \ket{\uparrow, \Uparrow } \), where in this notation, the first spin represents the electron spin, and the second one the nuclear spin. To see the nontrivial phase in the rotor Jackiw-Rebbi model, we prepare the state at each site in the $\ket{\downarrow, \Downarrow}$ state, instead of the ground state, that is $\ket{\downarrow \Uparrow} $ at all sites.
The electrons can be polarized to all-$\ket{\Downarrow}$ state using the strong magnetic field.
To achieve the desired spin configuration of nuclear spins, one can use the method that has been widely used for high-fidelity nuclear-spin control \cite{pla2013high}. In this method, the nuclear spin can be read via the electron-spin-resonance (ESR) method with high fidelity. Depending on the orientation of the nuclear spin, the external magnetic field experienced by the electron will vary. This variation leads to different resonance frequencies, $\nu_{e}^{\pm}=\gamma_e B\pm g/2$, for the two distinct nuclear-spin orientations. As a result, it allows for the detection of nuclear-spin states, $\ket{\Uparrow}$ or $\ket{\Downarrow }$. After reading the nuclear spin with this method, if the spin is already polarized as $\ket{\Downarrow}$, no operation is needed, otherwise the nuclear spin on that site can be polarized to $\ket{\Downarrow}$ using NMR fields or using the electric-dipole spin resonance (EDSR) technique of Ref.~\cite{Savytskyy23}. This polarization process based on measurement and feedback would need to be repeated individually on each site or parallelized across the array. Alternatively, a more easily scalable approach for large systems is to place a single electron on each dopant site and use dynamical nuclear polarization (DNP) to simultaneously pump the nuclear spins to a polarized state~\cite{Jarvinen14,Savytskyy23}. 

To engineer the right conditions for the rotor Jackiw-Rebbi model, after polarizing the nuclear spins, we need to change the electron occupation so that there is close to $\frac{1}{2}$ an electron for each dopant site. By choosing the detuning of the nuclear NMR field, one can also ensure that the correct sign of the $h_z$ term is obtained to realize the symmetry-broken phase. Once these conditions are satisfied, we envision that the ground state could be adiabatically prepared by starting from a state with a large chemical-potential difference between each site and then slowly tuning these differences to zero. This can be achieved by controlling the voltages of the gates at each site. Alternatively, it would also be possible to start in a state with zero chemical-potential differences and adiabatically tune the sign of the NMR detuning from positive to negative values. 

\subsection{
Accessible parameter regimes}\label{Sec:Experiment}
In order to lay the groundwork for analyzing nonperturbative phenomena within (2+1)D systems, selecting simulation parameters that align with the experimental capability is critical. Moreover, the optimal set of parameters needs to be meticulously determined. In Ref.~\cite{wang2022experimental}, the lattice constant, $a$, was explored within a range from $10~\text{nm}$ to $4~\text{nm}$. We have chosen $a=4.7~\text{nm}$ for most of our simulations. Due to the oscillatory behavior of the electron's wave function and the exponential dependence of the tunneling rate on the lattice constant, an accurate value of the lattice constant is crucial in obtaining definitive results. 
Such a high level of control on the value of $a$ has, nonetheless, been recently achieved~\cite{wang2022experimental}. With no adverse effects of disorder on resonant tunneling in the two-dimensional arrays, the dopant platform is favorable for studying models in (2+1)D~\cite{le2017extended}. 

The hyperfine coupling, which reflects the interaction of electron spin with nuclear spin, is almost a steady parameter (or can be slightly adjusted by applying a strong external electric field). Thus, the main macroscopic variable that allows one to feasibly access the full range of the phase diagram is the ratio of the hyperfine constant to the tunneling rate, $g/t$, while the external magnetic field can be varied from zero to a higher value. In the following, we discuss the experimental aspects concerning $t$ and $g$.

Let us denote the tunneling rates between sites $i$ and $j$ as $t_{ij}$. For an electron at site $i$ with wave function $|\psi_i\rangle$, we characterize the Coulomb potential, incorporating a screening factor $\lambda$, by the formula $U_i=- {V_0}e^{-\lambda |\bm{r}-\bm{r}_i|}/|{\bm{r}-\bm{r}_i|}$. We further define the overlap of wave functions between sites $i$ and $j$ as $S_{ij} = \langle \psi_i |\psi_j\rangle$, and the potential energy due to the interaction at site $j$ on an electron at site $i$ as $V_{ij} = \langle \psi_i| U_j |\psi_i\rangle$. Further define $V_{ij}' = \langle \psi_i| U_j |\psi_j\rangle$. With these definitions, one can calculate the tunneling rate $t_{ij}$ using the expression $t_{ij} = ({S_{ij}V_{ij}-V'_{ij}})/({1-S_{ij}^2})$, as detailed in Refs.~\cite{hu2005charge,le2017extended}. Given the wave function's exponential decay, it is reasonable to consider $t_{ij}$ as approximately zero for sites that are more than two lattice constants apart. For simplicity, we assumed equal spacing along both directions of the lattice, which simplifies $t_{ij}$ to $t$. In our simulations with spacing $a=4.7~\text{nm}$, we have taken $t=7.5~\rm{meV}$, consistent with Ref.~\cite{le2017extended}.

The tunneling rate and its profile are strongly influenced by silicon orientation. Silicon wafer orientation refers to the crystallographic directions and planes of the silicon crystal lattice. Each number in these orientations represents a Miller Index, which is a notation system in crystallography used to describe the orientation of planes within a crystal lattice. Among the three main orientations $[100]$, $[111]$, and $[110]$, the last is more favorable for experimental fabrications.
To estimate the profile of tunneling as a function of the distance between two adjacent sites, one can use the effective-mass theory, as outlined in Refs. \cite{gamble2015multivalley,dusko2014splitting,kohn1955theory} and summarized in Appendix~\ref{app:orientations}.

The hyperfine coupling, $g$, and the hyperfine splitting energy of phosphorus are contingent upon the specific isotope of phosphorus. For instance, \(^{31}\text{P}\), the most prevalent isotope, has a hyperfine splitting energy approximately in the range of \(114.3\) MHz to \(117\) MHz. 
This coupling can be affected by the Stark shift, 
which varies as \(g(\mathcal{E}) = g(0)(1 + 2.8 \times 10^{-3}\mathcal{E}^2)\) with $g(0)=0.48~\mu\text{eV}$ for external electric field, $\mathcal{E}$~\cite{feher1959electron}.
It can also be affected by other effects such as the engineering of the spin-orbit coupling for a given $\mathcal{E}$.

\begin{figure}[t!]
\includegraphics[width=0.5\textwidth]{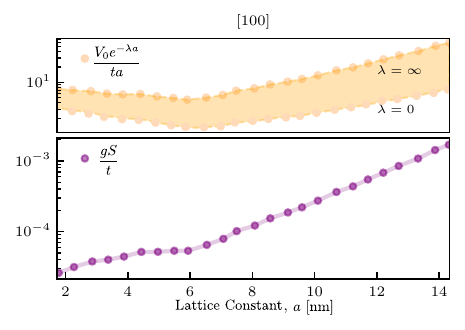}
\centering
\caption{\label{Fig:100}\textbf{Effect of lattice constant.}
The influence of lattice constant, \( a \), on the tunneling coupling, as well as the ratios \( g/t \) and \( h_z/t \) for fixed values of \( g =0.48~\mu\text{eV}\) and \( h_z=2.5~\text{T} \), is depicted. The top panel shows the ratio of the Coulomb potential coefficient, $V_0$, compared to the tunnel coupling across the full range of screening, $\lambda$. The bottom panel shows values of $gS/t$ for fixed $g=0.48~\mu \text{eV}$ as a function of the lattice constant.
}  
\end{figure}

Having access to the full range of the phase space, we need to make sure $h_z/t$ is of the same order as $gS/t$. The feasible region for observing the N\'eel phase in terms of lattice constant is depicted in Fig.~\ref{Fig:100} for the [100] silicon orientation. For the sake of completeness, other orientations are presented in Appendix~\ref{app:orientations}, Figs.~\ref{Figure:110}, and \ref{Figure:111}. As observed in Fig.~\ref{Fig:100}, for any given lattice constant, there always exists a narrow corridor of external magnetic fields that enable the manifestation of the nontrivial phase in the experimental setup. This corridor is wider in the [111] orientation as shown in Appendix~\ref{app:orientations}. 
\begin{figure}[!]
\includegraphics[width=0.5\textwidth]{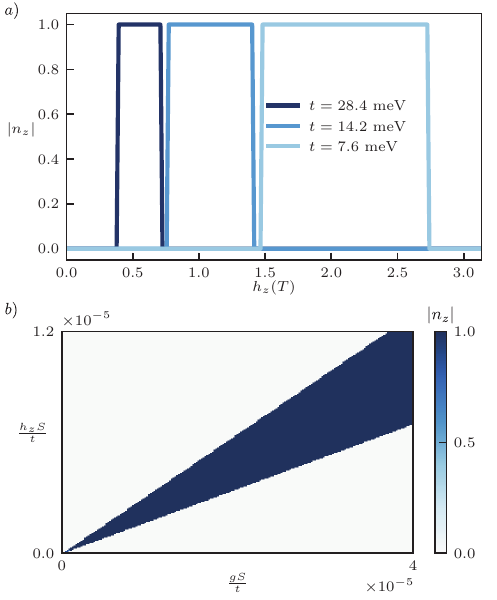}
\centering
\caption{\label{Fig:corner}\textbf{External magnetic field as a macroscopic control.}
a) The system is expected to transition through different phases corresponding to varying levels of the (longitudinal) magnetic field for three different values of tunneling coupling. The transition region from the trivial phase ($|n_z|=0$) to the N\'eel ordered phase ($|n_z|=1$) is illustrated for a $2 \times 2$ dopant array using the ED method.
b) The phase diagram of the $2 \times 2$ dopant array, with the region $gS/t$ coupling feasible in an experimental setup, has been simulated using the ED method. For this, the tunneling coupling is set at $t=7.5~\text{meV}$ and $h_xS/t=10^{-7}$.
}
\end{figure}

Variation of $gS/t$ during the experiment can be challenging. A more feasible approach is to keep $gS/t$ fixed and vary the external magnetic field as a macroscopic parameter, which can still change the system's phase. In this scenario, we are limited to exploring one slice of the 2D phase diagrams studied in Sec.~\ref{Sec:1+1d}. Due to the small value of $gS/t$ within the accessible range of parameters, $gS/t \sim 3.2 \times 10^{-5}$, experimental probing occurs at the corner of the phase diagram. Simulating this corner with the HF method is unreliable. However, the ED method for small arrays, as shown in Fig. \ref{Fig:corner}, indicates that as long as $|h_x| \ll |h_z|$, in a region where $h_zS/t$ is comparable to $gS/t$, the N\'eel phase is expected to be observed. This is indicative of the robustness of nonperturbative phenomena in this system. The numerical simulation in Fig.~\ref{Fig:corner}(a) shows that for a lattice constant of $a=4.7~\text{nm}$, we expect to observe the N\'eel phase when the magnetic field is around $1.5-2.5$ T. This range is feasible in the current experimental setups.

\subsection{Probing the system}
In this section, we continue our experimental feasibility study by exploring methods to detect the system and identify possible observables.
The dopant array is surrounded by electrons from outside of array, which allows electrons to tunnel to the island of the array via source and drain probes. Due to this tunneling capability of the electrons and the system's finite temperature, the total electron count is not fixed.

\begin{figure}[!]
\includegraphics[width=0.5\textwidth]{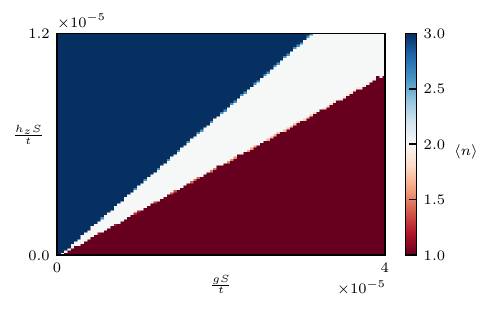}
\centering
\caption{\label{fig:charge}\textbf{Charge-occupation profile as a probe.} The variations in charge-occupation profiles, detected using a charge sensor, reflect the different phases in the $2 \times 2$ square array.
The pronounced differences in the charge profiles, particularly in the nontrivial phase, provide evidence for the detection of phase transitions. The charges are calculated by tracing the electrons' density matrix determined at a finite temperature of $T=10~ \text{mK}$ using the grand canonical ED method. Here,  $\langle n \rangle = \sum_i \langle c_i^\dagger c_i \rangle_\beta $. Other parameters used for this simulation are $t=7.5~\text{meV}$, $h_xS/t=0.01$, and $\mu/t=0.91$. 
} 
\label{Fig:charge-2d}
\end{figure}

One method for detecting the phase of a system involves using a global charge sensor. Such a sensor, which could be a quantum dot situated nearby or a proximal metallic gate, is acutely responsive to changes in the surrounding charge environment. By meticulously tracking variations in the conductance of this sensor, it becomes feasible to discern the addition or subtraction of a single electron from the dopant array. This ability allows for the prompt and precise detection of both the total charge and its transitions. As demonstrated in Fig.~\ref{fig:charge}, our numerical simulations for the (2+1)D system on a $2 \times 2$ square array reveal that the charge profiles, with respect to variations in $g$ and $h_z$, are discernible, effectively differentiating between the trivial and nontrivial phases. This method serves as a valuable tool for distinguishing between the phases, thereby enhancing our understanding of the system's phase transition characteristics. A similar study for (1+1)D dimensions has been demonstrated in Appendix~\ref{app:supp_plots}, Fig.~\ref{fig:charge_1d}.

Another method to analyze the state of the system involves investigating the linear conductance, denoted by \(G\), in relation to the chemical potential, denoted by \(\mu\). Considering a two-dimensional array where the source and drain interact primarily with the array's leftmost and rightmost columns---and their interaction with the inner parts of the array decreases exponentially---we can illustrate how the dopant array interacts with the leads. This method has been reviewed in Appendix~\ref{appendix:linear_conduct}. Our study shows that the pattern of linear conductance can differ depending on which of the two distinct phases the system is in. Nonetheless, the phase-discerning potential of this method is not as prominent as the charge-sensing method.

\section{Conclusion and outlook}
\noindent
In this study, we proposed utilizing a solid-state framework, specifically a donor array in silicon, as a native fermionic-bosonic analog quantum simulator. The primary objective was to explore this framework's applicability in studying relativistic phenomena in a quantum field theory, such as dynamical mass generation and confinement-deconfinement phase transitions.

Our study focused on the lattice representation of the rotor Jackiw-Rebbi model. Using numerical mean-field techniques, we explored the model's behavior in both (1+1)D and (2+1)D systems. Notably, for the rotor Jackiw-Rebbi model on a spatial square lattice, we discerned a nontrivial phase akin to its (1+1)D counterpart. In this phase, fermions acquire dynamical mass within the lattice framework, facilitated by an antiferromagnetic order of dopant spin acting as a bosonic field interacting with the fermions. The (1+1)D version of the model showcases the potential to enter a phase wherein charge fractionalization occurs, yielding transitions between confined and deconfined phases involving quark-like quasi-particles. Additionally, thermal effects or an external (transverse) magnetic field could reestablish chiral symmetry by dynamical instabilities akin to those observed in spin models~\cite{RodriguezNieva2022}. Extending our (2+1)D model to the honeycomb lattice reveals the presence of a similar nontrivial N\'eel-order phase. Moreover, phenomena analogous to those observed near the confinement-deconfinement transition can be traced in the creation of a superconductivity phase on a two-dimensional array over a certain range of model parameters, as studied in previous work~\cite{PhysRevResearch.5.L012009}.

Our research further elucidates the experimental feasibility of our proposal, paving the way for potential laboratory realizations and verifications. In relation to existing works, we juxtaposed our framework with proposals in the realm of cold atoms. This comparison emphasized the potential advantage of the solid-state platform of this work, particularly when transitioning to (2+1)D with native fermionic-bosonic simulators. Looking at the broader picture, our approach heralds the potential for empirical observations of confinement-deconfinement phase transitions, as well as the restoration of chiral symmetry at finite temperatures and densities, in a quantum field theory. 

While the model studied in this work presents a streamlined approach to analog quantum simulations, it is essential to recognize its differences compared with lattice gauge theories (LGTs). Unlike LGTs, the rotor Jackiw-Rebbi model does not exhibit gauge invariance, simplifying its implementation within experimental setups. This model, nonetheless, offers a more accessible introduction to analog quantum simulations of quantum field theories, and opens up opportunities to delve into the nonperturbative properties of quantum fields. As we look ahead, our ambition is to develop quantum-control techniques that enforce stringent gauge constraints on the degrees of freedom, making them more suitable for simulating LGTs. This will likely require time-dependent (floquet) engineering and novel approaches to develop a more intricate relationship between nuclear spins and fermions.

In summary, by harnessing engineered artificial lattices using solid-state technologies, we envision significant advancements in quantum simulations of quantum field theories, leading to deeper insights into strongly interacting systems in nuclear and high-energy physics.

\section*{Acknowledgments}
\noindent
We acknowledge discussions with Vinay Vikramaditya on the outlook of this work for simulating lattice gauge theories. This work was supported by the Department of Energy (DOE), Office of Science, Office of Nuclear Physics, via the program on Quantum Horizons: QIS Research and Innovation for Nuclear Science (award no. DE-SC0023710), the National Science Foundation's Quantum Leap Challenge Institute in Robust Quantum Simulation (award no. OMA-2120757), DOE's
Quantum Systems Accelerator program (award no. DE-AC02-05CH11231), and the Minta Martin and Simons Foundation.

\bibliography{main}
\newpage
\appendix

\section{Continuum Limit and Effect of Boson-Fermion Interactions}\label{app:CL}
\noindent
The connection between lattice Hamiltonians to relativistic quantum field theories is established in a standard way by resorting to the low-energy limit of the former. In this appendix, we examine the low-energy limit of the Hamiltonian in Eq.~\eqref{Eq:lattice_model}, first on the square lattice and then on the honeycomb lattice. The connection to Dirac fermions of the continuum theory is made clear in the latter case. The case of the (1+1)D lattice is analyzed in detail in prior work, including in Ref.~\cite{gonzalez2020rotor}.

\subsection{Square lattice: Fermi surface coupled to nuclear spins}
\begin{figure}[t!]
\includegraphics[width=0.4\textwidth]{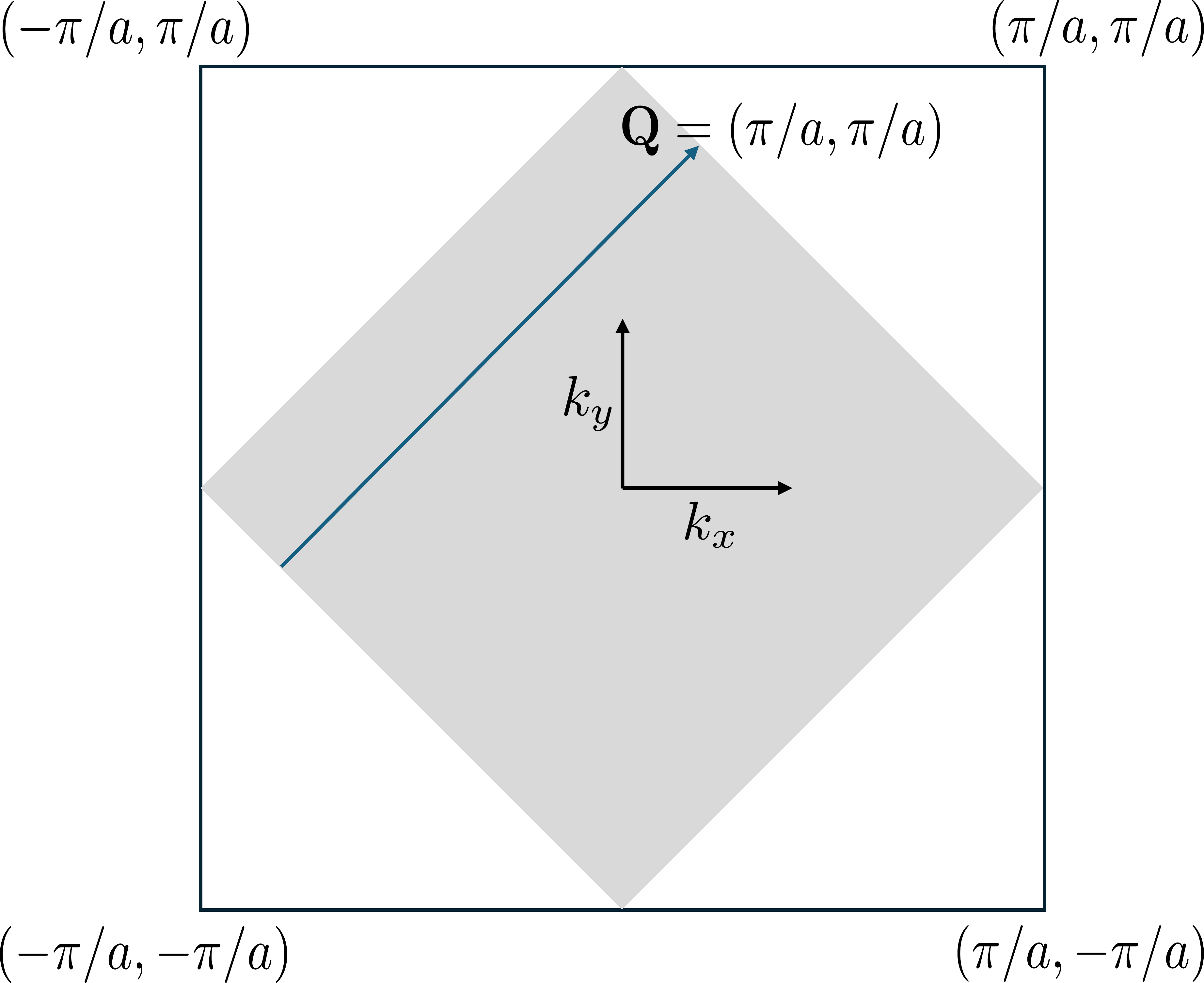}
\caption{\label{Fig:squareFS} 
\textbf{Tight-binding model at half-filling on a square lattice.} The Brillouin zone (square) and the Fermi surface of the square-lattice tight-binding model at half-filling (grey diamond) are depicted. The nesting vector $\bm{Q}=(\pi/a, \pi/a)$ is also shown, see the text for details.}
\end{figure}

First, we consider the free (kinetic) part of Eq.~\eqref{Eq:lattice_model} on a square lattice, which represents a tight-binding fermion model,
\begin{align}\label{Eq:free_model}
\begin{split}
    H_{\text{k}} = & -\sum_{\langle i, j \rangle} t\left(c_i^{\dagger} c_j + \text{h.c.}\right) - \mu \sum_i n_i.
\end{split}
\end{align}
We assume the electron chemical potential is uniform throughout the lattice and further restrict it to $\mu = 0$, i.e., the half-filling limit. Let us define the momentum-space fermionic operator,
\begin{equation}
    c_{\bm{k}} = \frac{1}{\sqrt{N}} \sum_i e^{-i \bm{k} \cdot \bm{r}_i} c_i,
\end{equation}
where $N$ is the number of lattice sites, $\bm{r}_i$ is the position vector associated with site $i$, and $\bm{k}$ is the momentum. In this basis,
\begin{equation}
    H_{\text{k}} = -2t \sum_{\bm{k}} \big[\cos (k_x a) + \cos (k_y a)\big] c^{\dagger}_{\bm{k}} c_{\bm{k}}.
\end{equation}
The ground state of $H_{\text{free}}$ is depicted in Fig.~\ref{Fig:squareFS}. All states within the diamond-shaped Fermi surface (shaded grey) are occupied.

Now consider the interaction of electrons with the nuclear spins in Eq.~(\ref{Eq:lattice_model}), which in momentum space reads: $-g\sum_{\bm{q},\bm{k}} c^{\dagger}_{\bm{k}+\bm{q}} c_{\bm{k}} I^z_{\bm{q}}$. Notice in Fig.~\ref{Fig:squareFS} that the Fermi surface has a nesting vector of $\bm{Q}=(\pi/a, \pi/a)$: for any point $\bm{k}$ that lies on the Fermi surface, $\bm{k}+\bm{Q}$ also lies on the Fermi surface. Thus, any bosonic mode of momentum $\bm{Q}$ that couples to the fermions, scatters an electron on the Fermi surface to a different point also on the Fermi surface, leaving its energy unchanged. Since we are interested in low-energy excitations of fermions around the Fermi surface, it is useful to focus on the bosonic modes centered around $\bm{q}=\bm{0}$ and $\bm{q}=\bm{Q}$. For us, the bosons in question are the nuclear spins. Then, $\bm{q}=0$ corresponds to uniform spin order and $\bm{q}=\bm{Q}=(\pi/a,\pi/a)$ corresponds to N\'eel order. This provides further motivation for our choice of decomposition of the nuclear spins into the N\'eel and uniform components in Eq.~\eqref{Eq:nspindecomp}. Let us assume $a^{2d}\bm{\ell}^2 \ll 1$ and decompose the unit N\'eel vector $\bm{\phi}$ as
\begin{equation}
    \bm{\phi}(x,y)=\phi_0 \hat{\bm{z}}+ \delta\bm{\phi}(x,y),
\end{equation}
where $ \phi_0 \hat{\bm{z}}\equiv\expval{\bm{\phi}}$ is constant over each lattice site, and $\delta\bm{\phi}\equiv \bm{\phi}-\expval{\bm{\phi}}$ is the fluctuating part of the N\'eel field. Here, $\hat{\bm{z}}$ is a unit vector in the internal (nuclear-spin) $z$-direction. So, 
\begin{align}
    \bm{I}(x,y)=
    S\left[ (-1)^{\frac{x}{a}+\frac{y}{a}}\left(\phi_0 \hat{\bm{z}}+\delta\bm{\phi}(x,y)\right)+a^2 \bm{\ell}(x,y)\right].
\end{align}
With this ansatz, consider the interaction between the fermions and nuclear spins. We see that when the average N\'eel field $\phi_0$ is nonzero, there is a term that couples every fermion at the Fermi surface to a different Fermion $\bm{Q}$ away in momentum space (Fig.~\ref{Fig:squareFS}). As a consequence, the degeneracy between two points on the Fermi surface that are separated by $\bm{Q}$ is split. Hence, a gap opens up along the Fermi surface, which means at half-filling, the occupied free-fermion state is gapped. This also results in the halving of the Brillouin zone down to the shaded region in Fig.~\ref{Fig:squareFS}.
\begin{figure}
\includegraphics[width=0.32\textwidth]{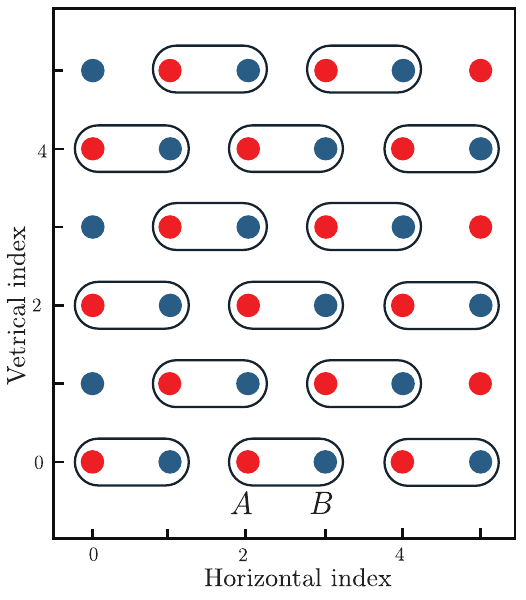}
\caption{\label{Fig:staggeredlattice}\textbf{Square lattice and N\'eel order.} Schematic of the putative lattice structure in the lattice fermionic-bosonic model when $\phi_0\neq 0$, i.e., when there is a nonzero N\'eel order. The two-site unit cell is marked. The sites labeled red (A sites) have nuclear spins pointing up (in the $\sigma^z$ basis) and those labeled blue (B sites) have nuclear spins pointing down.
}
\end{figure}
When $\phi_0\neq 0$, the new unit cell has a two-site A-B basis shown in Fig.~\ref{Fig:staggeredlattice}. Another symmetry that is broken is the reflection symmetry around lines parallel to the $x$ or $y$ axes that pass through points $\bm{R}+(\tfrac{a}{2},\tfrac{a}{2})$ for a site $\bm{R}$ belonging to the original square lattice. 

Let us now define
\begin{align}
    \label{eq:c-kA}
    c_{\bm{k},A}&=\sqrt{\frac{2}{N}}\sum_{i\in A}c_i e^{-i\bm{k}\cdot \bm{r}_i},\\
    \label{eq:c-kB}
    c_{\bm{k},B}&=\sqrt{\frac{2}{N}}\sum_{i\in B}c_i e^{-i\bm{k}\cdot \bm{r}_i},\\
    \label{eq:Psi-AB}
    \Psi_{\bm{k}}&\equiv \begin{pmatrix}c_{\bm{k},A}&  c_{\bm{k},B}\end{pmatrix}^T,
\end{align}
and further do a change of variables from $(k_x,k_y)$ to $(k_+,k_-)$ where $k_{\pm}=\tfrac{1}{\sqrt{2}}(k_x\pm k_y)$. The new Brillouin zone, i.e., the shaded region in Fig.~\ref{Fig:squareFS}, is given by 
\begin{equation}\label{Eq:BZdef}
    -\frac{\pi}{\sqrt{2}a}<k_{\pm}\leq \frac{\pi}{\sqrt{2}a}.
\end{equation}
In this notation, the fermionic Hamiltonian $H_f$ obtained by retaining only the terms involving the fermions in Eq.~\eqref{Eq:lattice_model} is given by
\begin{equation}\label{Eq:HFermion}
\begin{split}
    H_f=&\sum_{\bm{k}\in \overline{\text{BZ}}}\Psi^{\dagger}_{\bm{k}}\left\{-4t\cos \left(\frac{k_+ a}{\sqrt{2}} \right)\cos \left(\frac{k_- a}{\sqrt{2}} \right)\tau^x\right.\\
    & \quad \quad \quad \quad -gS\phi_0 \tau^z \biggr\}\Psi_{\bm{k}}\\
    &-gS\sum_{\bm{k},\bm{q}}\Psi^{\dagger}_{\bm{k}+\bm{q}}\left[\delta\phi^z_{\bm{q}}\,\tau^z +\ell^z_{\bm{q}}\right]\Psi_{\bm{k}}.
\end{split}
\end{equation}
Here, $\overline{\text{BZ}}$ denotes the reduced Brillouin zone defined in Eq.~\eqref{Eq:BZdef} and $\tau^i$ denote Pauli matrices in the A-B-site indices. The spectrum of the part of $H_f$ without the fluctuation terms of the last line is thus given by
\begin{equation}
\begin{split}
    \varepsilon_{f}(\bm{k})\\&=\pm\left\{\left(gS\phi_0\right)^2+16t^2 \cos^2 \left(\frac{k_+ a}{\sqrt{2}}\right)\cos^2 \left(\frac{k_- a}{\sqrt{2}}\right)\right\}^{1/2}.
    \end{split}
\end{equation}
We see that the higher-energy band has minimum energy along the boundary of the shaded Brillouin zone in Fig.~\ref{Fig:squareFS}, i.e. along $k_+=\pm \tfrac{\pi}{\sqrt{2}a}$, and separately along $k_-=\pm \tfrac{\pi}{\sqrt{2}a}$. Here, the minimum energy fermionic excitations occur along an entire locus of points in momentum space, as opposed to a single momentum, say $\bm{k}=\bm{0}$. Hence, this theory does not map to a (discretized) Dirac theory. Nonetheless, it exhibits the dynamical mass generation explored in this work.

\subsection{Honeycomb lattice: Dirac Fermions coupled to nuclear spins}
\begin{figure}[t!]
\includegraphics[width=0.5\textwidth]{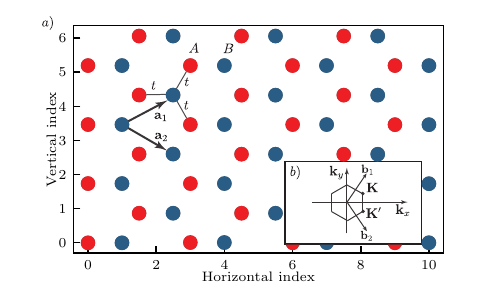}
\centering
\caption{\label{Fig:hex} \textbf{Tight-binding model at half-filling on a honeycomb lattice.} 
a) The interactions in a honeycomb lattice can be described as the interaction of each sublattice point, \( A \) and \( B \), with its three nearest neighbors, characterized by strength $t$. Basis vectors $\bm{a}_{1,2}$ are also shown. b) The Brillouin zone and the nodes, $\bm{K}$ and $\bm{K}'$, as discussed in the main text, along with the basic vectors on the reciprocal lattice.}  
\end{figure}
Let us revisit the interactions between fermions and nuclear spins on the honeycomb lattice, which readily reveal a Dirac-fermion property. To see how, we briefly review the tight-binding model of fermions on a honeycomb lattice. For this lattice array, one may choose basis vectors $\bm{a}_1 = \frac{a}{2}(3, \sqrt{3})$ and $\bm{a}_2 = \frac{a}{2}(3, -\sqrt{3})$, and the basis vector in the reciprocal lattice $\bm{b}_{1}=\frac{2\pi}{3a}(1,\sqrt{3}), \bm{b}_{2}=\frac{2\pi}{3a}(1,-\sqrt{3}) $ such that $\bm{b}_j\cdot \bm{a}_i=2\pi \delta_{i,j}$. Depending on the site, a fermion may hop forward along the $x$ axis, or hop along the $\bm{a}_1$ and $\bm{a}_2$ vectors. Therefore, it is useful to define two sublattices A and B, as shown in Fig.~\ref{Fig:hex}, each containing one type of the sites. Now upon defining $c_{\bm{k},A}$, $c_{\bm{k},B}$, and $\Psi_{\bm{k}}$ as in Eqs.~(\ref{eq:c-kA})-(\ref{eq:Psi-AB}), the fermion kinetic Hamiltonian can be written as:
\begin{equation}
H_\text{k}= \sum_{\bm{k}} \Psi_{\bm{k}}^\dagger h(\bm{k}) \Psi_{\bm{k}}
\end{equation}
with
\begin{align}
h(\bm{k})=&-t \, e^{ik_xa}\times
\nonumber\\
&\begin{pmatrix}
0 & e^{-i \bm{k} \cdot \bm{a}_1} + e^{-i \bm{k} \cdot \bm{a}_2} + 1 \\
e^{i\bm{k} \cdot \bm{a}_1} + e^{i \bm{k} \cdot \bm{a}_2} + 1 & 0 
\end{pmatrix}.
\end{align}
Let us now expand this Hamiltonian around the points of the Brillouin zone, $\bm{K} = \frac{2\pi}{3a}(1, \tfrac{1}{\sqrt{3}})$ and $\bm{K}' = \frac{2\pi}{3a}(1, -\tfrac{1}{\sqrt{3}})$ (which is equivalent to $-\bm{K}$ up to a reciprocal lattice vector). For values $\bm{k} = \pm \bm{K} + \bm{\kappa}$ with small $|\bm{\kappa}|$, this leads to 
\begin{align}
h(\pm\bm{K} + \bm{\kappa}) &= 
-v_F (\kappa_x  {\sigma}_y \mp \kappa_y \sigma_x )
\nonumber\\
&=-v_F (\mp \kappa_x i\gamma^2 \pm \kappa_y i\gamma^1),
\end{align}
at linear order in $\bm{\kappa}$ and upon a rescaling of the fermion fields with a phase. Here, we have defined the $\gamma$ matrices as $\gamma^0=\sigma^z$, $\gamma^1=i\sigma^x$, and $\gamma^2=\pm i\sigma^y$. Now note that $\gamma^0 h(\pm\bm{K}+\bm{\kappa})\gamma^0=-v_f(\kappa_x  \gamma^0 \gamma^1 + \kappa_y \gamma^0 \gamma^2)$. For this reason, we can define $\Psi$ as done in Eq.~\eqref{eq:defpsi1}, which is the transformed version of the original $\Psi_{\bm{k}}$, i.e., $\Psi_{\bm{k}} \to \gamma^0 \Psi_{\bm{k}}$.
Comparing this behavior with that of Dirac fermions, i.e., the kinetic term in the Hamiltonian in Eq.~(\ref{eq:H-RJR-d+1}), we observe that the behavior of electrons in the honeycomb lattice near the nodes identified can be approximated by massless Dirac fermions in (2+1)D with an effective speed of light $c=v_F = \frac{3ta}{2}$. Note that in odd spacetime dimensions, there are two irreducible representations for the Clifford algebra which are not related by a similarity transformation~\cite{park2022lecture}, hence the two sets of $\gamma$ matrices introduced above. Alternatively, one may work with one set of $\gamma$ matrices and still recover the Dirac equation of motion by expanding around both Dirac points $\pm \bm{K}$. This requires different redefinitions of $\Psi_{\bm{k}}$ at each Dirac point, which can be obtained straightforwardly.

Finally, upon coupling these fermions to bosons, and considering a phase with a nonzero $\phi_0$ (N\'eel order in the rotor picture), the reflection symmetry along e.g., lines parallel to the $y$ direction passing through hexagon centers gets broken, corresponding to spontaneous breaking of the $\Psi(x,y) \to \gamma^1 \Psi(-x,y)$ symmetry in the original Dirac Hamiltonian, hence a fermion mass gets generated dynamically.

\section{Supplementary Plots}\label{app:supp_plots}
\noindent
A number of supplemental results were left out from the main text and will be presented here, along with accompanying discussions. These include the effect of the chemical potential on the phase diagram (of a square array), a discussion of correlations in both the trivial and nontrivial phases, and the charge-occupation profile for a one-dimensional array.
\begin{figure*}
\includegraphics[width=\textwidth]{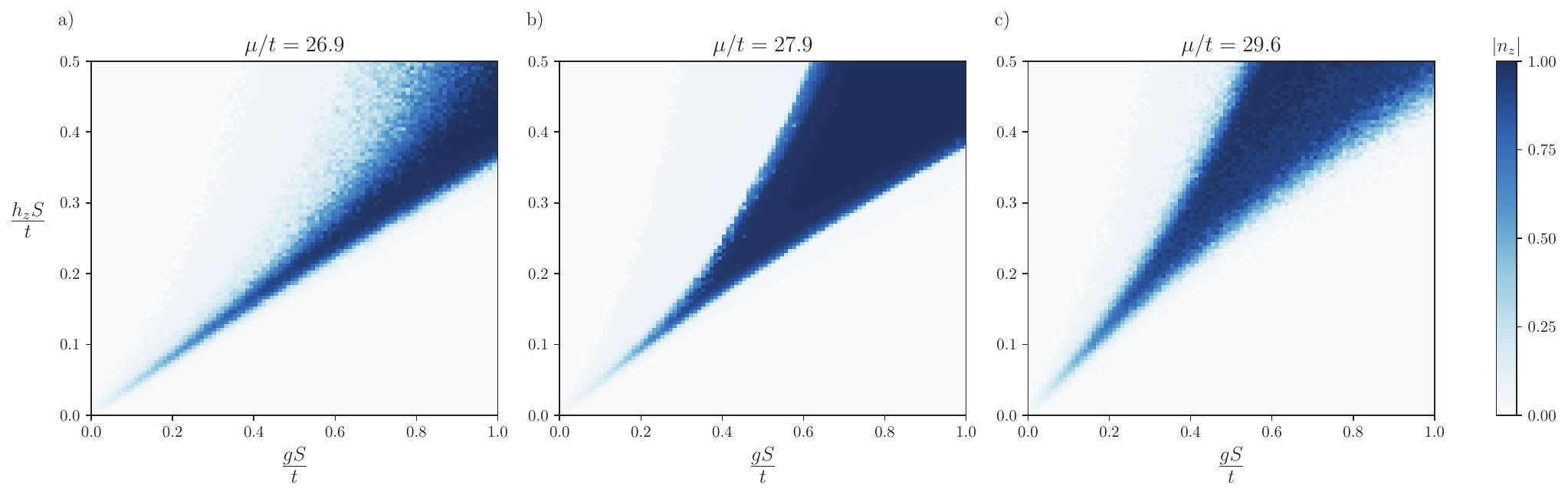}
\caption{\label{Fig:chemical_potential_effect} \textbf{Effect of chemical potential on the phase diagram.} This figure displays the phase diagram of a (2+1)D system with size of $N=10\times 10$ at half filling. $n_z$ denotes the order parameter defined in Eq.~\eqref{Eq:nz-def}. The diagram is rendered for three different chemical potentials, $\mu$, each normalized with the tunneling coupling $t=7.5~\text{meV}$. The inverse temperature for all subplots is uniformly fixed at $\beta=100$ and $h_xS/t = 0.01$.}
\end{figure*}
%

\subsection{Effect of chemical potential}
To examine the influence of the chemical potential on the system's phase, we showcase the phase diagram for three different chemical potentials in Fig.~\ref{Fig:chemical_potential_effect}. The chemical potential can be adjusted to regulate the electron transport from the Fermi sea (i.e., the surrounding probes) to the dopant array. Thus, to achieve a half-filling state, it is feasible to adjust the appropriate exterior chemical potential. This is particularly important in the finite-temperature scenario, where our control is limited to the expected value of the total number of electrons on the array.

\subsection{Trivial phase}\label{subseq:LL}
\begin{figure}
\includegraphics[width=0.5\textwidth]{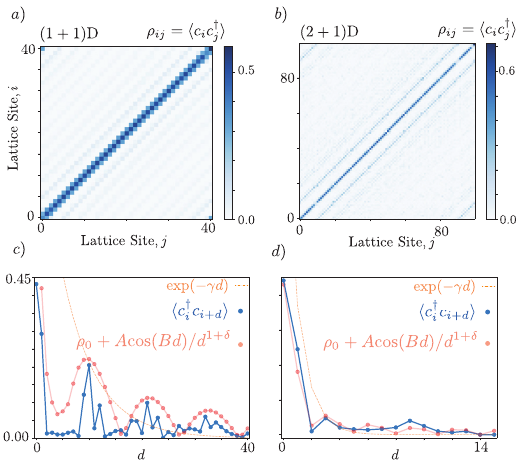}
\caption{\label{Fig:LL} \textbf{Correlation function and fermion behavior in the trivial phase.} a) and b) The single-particle density matrix of the fermions in their trivial phase illustrates behavior akin to a Luttinger Liquid in (1+1)D. c) The correlation function $\langle c_{i}^\dagger c_{i+d} \rangle$ is plotted as a function of $d$ for a fixed position $i=N/2$. The correlation function has been tested against two analytical functions suggested by theoretical predictions, see Sec.~\ref{subseq:LL} for more details. One behavior is exponential decay characterized by the parameter \(\gamma\), while the other is oscillatory decay characterized by the parameters \(\rho_0\), \(A\), $B$, and \(\delta\).  These parameters are obtained by minimizing the distance (mean square error) between the data and the analytical function. 
d) Analogous calculation for the (2+1)D scenario with size $N=10 \times 10$. The lattice indices in the (2+1)D case are defined by row-major order.}
\end{figure}

In the main text, our focus was on the nontrivial phase in which both the dynamical mass and N\'eel order are evident. For a comprehensive understanding, 
we briefly discuss the contrasting phase named `trivial'. In the (1+1)D context, this phase corresponds to the Luttinger-Liquid (LL) phase. 
When plotting the correlation function, 
\( \langle c_i^\dagger c_{i+d}\rangle \), 
against $d$ (the separation between two sites), an oscillatory pattern emerges. 
As illustrated in Fig.~\ref{Fig:LL}, this contrasts with an exponential trend 
and is consistent with the characteristic charge-density wave (CDW) 
observed in (1+1)D systems. 
The precise form of the correlation depends on the specific model and boundary conditions. For finite-temperature scenarios, this correlation can be approximated as:
\( \langle c_i^\dagger c_{i+d}\rangle \sim \rho_0 + A\cos(B d)/d^{1+{\delta}} + \cdots \)
where $\rho_0$, $A$, $B$, and $\delta$ are functions of the model's parameters, such as temperature and Fermi momentum \cite{fradkin2013field}. In the (2+1)D context, the theory of Luttinger Liquid cannot be extended straightforwardly. However, an oscillatory pattern remains evident, as shown in Fig.~\ref{Fig:LL}(b) and (d).

(Thermal) correlation functions, as well as total fermion occupations, are also plotted for the two-dimensional array for the nontrivial phase in Fig.~\ref{Fig:tr_rho,tr_K}. As explained in Sec.~\ref{Sec:dyn-mass-2+1d}, these results indicate that the pairing mechanism due to only Coulomb potential is unlikely to explain the nontrivial phase observed in this model. However, due to the limitations of the HF method, it cannot address other contributions to pairing, such as those caused by a combination of the Coulomb potential and electron-spin interactions. We leave these aspects to future studies.

\subsection{Charge-occupation profile in (1+1)D}
\begin{figure}[t!]
\includegraphics[width=0.5\textwidth]{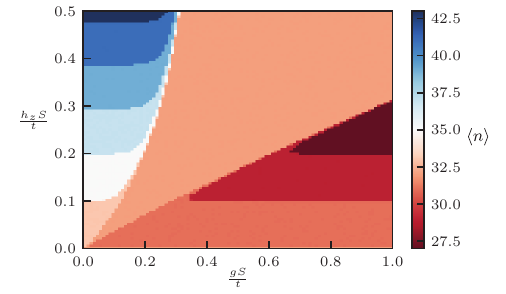}
\centering
\caption{\label{fig:charge_1d}\textbf{Charge-occupation profile for a (1+1)D system.} Similar to Fig.~\ref{fig:charge}, the variations in charge-occupation profiles have been simulated for the $N=64$ one-dimensional array. The charges are simulated by tracing the single-particle density matrix determined at a finite temperature of $T=10~\text{mK}$. Here, $\langle n \rangle=\sum_i \langle c_i^\dagger c_i \rangle_\beta $. The tunneling coupling is set at $t=7.5 \text{meV}$, $\mu/t=11.9$, $h_x S/t=0.01$, and periodic boundary condition are used. The calculations are performed using the FTHFB method. 
}  
\end{figure}

In alignment with Fig.~\ref{Fig:charge-2d} and the discussion in the experimental section \ref{sec:experimental_probing}, we exhibit the outcomes of the simulations for the charge-occupation profile of the (1+1)D system with $N=43$ at a finite temperature in Fig.~\ref{fig:charge_1d}. As shown, while the nontrivial phase exhibits a wide range of stable charge occupations across the phase diagram, the trivial phase is more layered, exhibiting different charges as the macroscopic parameters are changed. This feature can be used to probe the two phases in the one-dimensional array.

\begin{figure}
\includegraphics[width=0.5\textwidth]{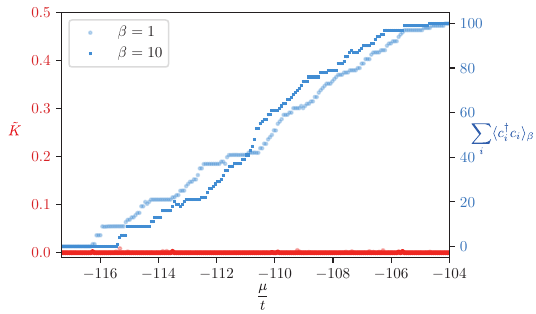}
\caption{\label{Fig:tr_rho,tr_K} \textbf{Correlation functions for a \(10 \times 10\) array at two inverse temperatures \(\beta\).} 
We show the total occupation number 
\(\sum_i \langle c_i^\dagger c_i \rangle_\beta\) (right axis) and the average pairing matrix 
\(\tilde{K} \coloneqq \sum_{i,j} {K_{ij} K_{ij}^*}/{N^2}\) with \(K_{ij} = \langle c_j c_i \rangle_\beta \). 
Here, the lattice constant is \(a = 4.7 \, \mathrm{nm}\), the tunneling coupling is \(t = 7.5 \, \mathrm{meV}\), 
the Coulomb potential strength is \(V_0S/(td) = 3.6\), and the transverse component of the external 
magnetic field is set to \(h_xS/t = 0.01\). As we can observe, \(\tilde{K}\) is almost zero for the entire range of chemical potential, suggesting that \(\langle c_i c_j \rangle\) is zero using the Hartree-Fock method. 
}
\end{figure}
%

\section{Finite-Temperature Hartee-Fock-Bogoluivov Method}\label{appendix_FTHFB}
\noindent
The Hartree-Fock method~\cite{hartree1928wave} constricts the wave function $|\psi\rangle$ to a specific form known as a Slater determinant, which aims to find the minimum energy that best approximates the system, thus finding a state as close as possible to the ground state. The Hartree-Fock-Bogoliubov (HFB) method~\cite{goodman1981finite} serves as a more generalized single-particle theory that builds upon the Hartree-Fock method.
In this method, one considers the actual Hamiltonian, $H$, defined in terms of fermionic operators ${c_i^\dagger,c_i}$, and approximates this using quasi-particle operators ${a_i^\dagger, a_i}$, yielding an approximate Hamiltonian $\tilde{H}=\sum_{i=1}^N \tilde{E}_i a_i^\dagger a_i$ with the approximate ground state $|\Phi\rangle$. In order to effectively apply the HFB method, one needs to expand the energy expectation value \(\frac{\langle \Phi |H|\Phi\rangle}{\langle \Phi| \Phi\rangle}\) up to the second order in terms of the largest energy scale in the Hamiltonian. In the Fermi-Hubbard model, this is normally the hopping terms, associated with the coupling \( t\).

The HFB method at finite temperature can be established by minimizing the grand canonical ensemble's free energy
\begin{equation}
    \Omega=E-T\mathcal{S}-\mu n.
\end{equation}
with $\mathcal{S}$ standing for the van-Neumann entropy and $n$ standing for electron occupation. 
We start with the transition from the fermionic operator basis, represented as $\{c\}$, to the quasi-particle basis, denoted as $\{a\}$:
\begin{align}
   & a_i^\dagger= \sum_j \left(U_{ij}c_j^\dagger +V_{ij}c_j\right),
   \nonumber\\
   &a_i=\sum_j \left(V_{ij}^* c_j^\dagger +U_{ij}^* c_j\right),
\end{align}
such that the matrices $U$ and $V$ satisfy the following set of identities
\begin{equation}
    U U^\dagger +V V^\dagger=1 , \,\,\,\, U V^T+V U^T=0.
\end{equation}
The single-particle density matrix, represented as $\rho_{ij}=\langle a_j^\dagger a_i\rangle$, can be rewritten  in the following manner:
\begin{equation}
\rho= U^T f U^{*}+V^\dagger (1-f)V,
\end{equation}
where $f(E)=(1+e^{E/(k_BT)})^{-1}$ is the Fermi-Dirac distribution. Likewise, the pairing matrix, denoted as $K_{ij}=\langle a_j a_i\rangle$, can be described as:
\begin{equation}
K=U^T f V^* +V^\dagger (1-f)U,
\end{equation}

In finite-temperature conditions, the vacuum state of quasi-particles no longer serves as a reference for the product state, causing Wick's theorem to be inapplicable. Despite this, ensemble averaging remains a valid technique for statistical analysis~\cite{goodman1981finite,abrikosov2012methods}.
In the case of a Hamiltonian featuring quartic interaction, the interaction component can be approximated as follows:
\begin{equation}
    \langle c_i^\dagger c_j^\dagger c_\ell c_k \rangle =\langle c_i^\dagger c_k \rangle \langle c_j^\dagger c_\ell\rangle -\langle c_i^\dagger c_\ell \rangle \langle c_j^\dagger c_k \rangle +\langle c_i^\dagger c_j^\dagger \rangle \langle c_\ell c_k \rangle.
\end{equation}

Leveraging the aforementioned relationship, we can express a general  Hamiltonian, which has up to quadratic interaction, with general tunneling matrix $t$, as:
\begin{align}
    \begin{split}
        H_{ij}&=t_{ij}+\Gamma_{ij}-\mu,\\
\Gamma_{ij}&=\sum_{i,j,k,\ell}\bar{V}_{ijk\ell}\rho_{k\ell},\\
\Delta_{ij}&=\frac{1}{2}\sum_{k,\ell}\bar{V}_{ijk\ell}K_{k\ell},
    \end{split}
\end{align}
and the energy can be described through the following expression
\begin{equation}
    E=\Tr[t\rho +\frac{1}{2}\Gamma \rho]-\Tr\left[\frac{1}{2}\Delta K^{*}\right].
\end{equation}

Similarly, the grand-canonical free energy can be expressed as
\begin{align}
    \begin{split}
        \Omega&=\sum_{i,j}(\Gamma-\mu)_{ij}\rho_{ji}+\frac{1}{2}\sum_{i,j,k,\ell}\bar{V}_{ijk\ell}\rho_{\ell j}\rho_{ki}\\
        &+\frac{1}{4}\sum_{i,j,k,\ell}\bar{V}_{ijk\ell}K_{ij}^{*}K_{k\ell}\\
        &+k_BT \sum_{i}\left[f_i \ln f_i +(1-f_i)\ln(1-f_i)\right].
    \end{split}
\end{align}

By setting the variation of the grand canonical potential, denoted as $\delta \Omega$, equal to zero at a given value of $\beta$, one can determine the ground state of the system
\begin{equation}
   \begin{pmatrix}
H & \Delta\\
-\Delta^* & -H^*
\end{pmatrix} 
\begin{pmatrix}
U_i\\
V_i
\end{pmatrix}
=E_i
\begin{pmatrix}
    U_i \\
    V_i
\end{pmatrix}.
\end{equation}

This set of equations is markedly nonconvex. Nonetheless, it is expected that with a suitable initial setup, one can achieve convergence to the global minimum with iterative methods. We have used this method in the main text and Appendices for the majority of our calculations of fermion densities and correlations. 

%

\section{Dopants in Silicon
\label{app:orientations}}
\noindent
In this section, we briefly outline the general description of an electron's wave function bound to a dopant in silicon and its connection to calculating the tunneling coupling of an electron from one dopant site to another.
Each donor has the ability to be host of one electron in its neutral state, referred to as \( D0 \), or two electrons when in its negatively charged state, termed \( D- \). The binding energy for an isolated neutral \( D0 \) center is approximately \( 45~\text{meV} \). In this study, our main focus is on the ground state of the donor in silicon, denoted as \( 1sA_1 \) in the literature.

The wave function of an electron can be written as an expansion of Bloch functions around the lower part of the conduction band. The conduction band of silicon can be characterized by six valleys located at 
\[
\bm{K} = 0.85 \times \frac{2\pi}{a_{\text{Si}}} \{[\pm 1,0,0],[0,\pm 1,0],[0,0,\pm 1] \}
\]
from the \(\Gamma\) point of the Brillouin zone. The \(\Gamma\) point represents the center of the Brillouin zone, where the wave vector \(\bm{k}\) is zero. It serves as a reference point for defining other points in the reciprocal space. The lattice constant of silicon, denoted as \(a_{\text{Si}}\), is approximately \(0.54 \text{ nm}\) ~\cite{calderon2009quantum}. The 6-fold expansion of the wave function around the sub-valley conduction ~\cite{seitz1957solid} can be written as:
\begin{equation}
    \psi_{A_1}(\bm{r})=\frac{1}{\sqrt{6}}\sum_{i\in\pm \{ x, y, z\}}F_{i}(\bm{r})\phi_{i}(\bm{r}).
    \label{eq:psi-A1}
\end{equation}
Here, \( F(\bm{r}) \) represents the slowly varying envelope function, while \( \phi(\bm{r}) \) denotes the Bloch function for each valley ~\cite{kohn1955theory}. These Bloch functions can be characterized by~\cite{gamble2015multivalley}:
\begin{equation}
\phi_i(\bm{r})=u_i(\bm{r})e^{i \bm{K_i}\cdot \bm{r}}.
\end{equation}
When multiple dopants are present, the interaction of electrons from different valleys, known as valley-orbital coupling, leads to the breaking of the 6-fold degeneracy. Typically, only two states, \(\pm z\), have lower energy. The wave function can be expressed as a superposition of these two components:
\begin{align}
\psi_{A_1} &= \frac{1}{\sqrt{2}}(\psi_{+z} + \psi_{-z})\\
&=\frac{1}{\sqrt{2}}F(\bm{r})[\phi_{+z}(\bm{r})\pm \phi_{-z}(\bm{r})].
\end{align}

The envelope function can be determined via the variational method. Following Ref.~\cite{gamble2015multivalley}, one can expand it over a finite set of orbital basis states: 
\begin{equation}
    F_i(\bm{r})=\sum_{\nu=1}^{N} A_{i,\nu}F_{i,\nu}(\bm{r})
\end{equation}
such that for example:
\begin{equation}
    F_{z,\nu}\sim x^{n_x}y^{n_y}z^{n_z}e^{-\alpha_\perp(x^2+y^2)}e^{-\alpha_\parallel z^2}\end{equation}
The parameters \(\{ n_x, n_y, n_z, \alpha_\perp, \alpha_\parallel \}\) need to be obtained numerically using the variational method.

To write the full Hamiltonian for the variational method, one can use the Shindo-Nara multi-valley effective-mass theory~\cite{shindo1976effective} to express the kinematic term as a combination of two terms with two masses. For example, in the \(+z\) direction:
\begin{equation}
    T_z=-\frac{\hbar^2}{2m_\perp}\left[\frac{\partial^2}{\partial x^2}+\frac{\partial^2}{\partial y^2}
    +\gamma\frac{\partial^2}{\partial z^2}\right],
\end{equation}
where \(\gamma \coloneqq m_{\perp}/m_\parallel\) with \(m_\perp = 0.19\,m\) and \(m_\parallel = \,0.91m\). The full Hamiltonian can then be written as:
\begin{equation}
    E F_{i}(\bm{r})=[\bm{T}_i+U(\bm{r})]F_i(\bm{r})+\sum_{\lambda \in \pm \{x,y,z\}}C_{\lambda i}(\bm{r})F_{i}(\bm{r}).
\end{equation}
Here, \( F_i \) was introduced after Eq.~\eqref{eq:psi-A1}, \( \bm{T}_i \) denotes the kinetic term that encompasses two effective masses, symbolized as \( m_\perp \) and \( m_\parallel \). The attractive binding potential $U(\bm{r})$ due to a donor in silicon is well approximated as a screened Coulomb potential at long distances, up to corrections that are discussed in Ref.~\cite{gamble2015multivalley}.  
\( C_{\lambda i}(\bm{r})\coloneqq \phi^{*}_\lambda (\bm{r})\phi_i(\bm{r})U(\bm{r}) \) represents the coupling between different valleys. Leveraging these mathematical constructs enables a method to find the envelope function, \( F(\bm{r}) \), and Bloch functions, $u_i(\bm{r})$, by minimizing the ground-state energy using the variational method. This procedure leads to a full description of the wave function in Eq.~\eqref{eq:psi-A1}. The tunneling coupling between dopant sites can then be calculated, as outlined in Section~\ref{Sec:Experiment}.

\begin{figure}
\includegraphics[width=0.5\textwidth]{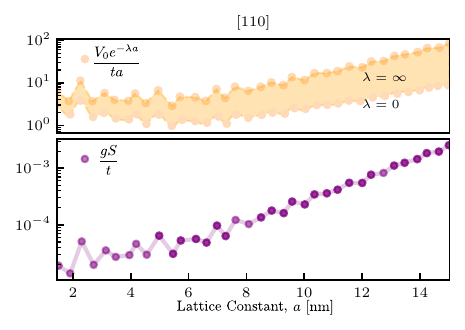}
\centering
\caption{\label{Figure:110}\textbf{Effect of lattice constant for [110] orientation.}
The effect of the lattice constant, $a$, on the tunneling coupling, $t$, is plotted for a $[110]$ silicon crystal orientation for fixed values of \( g =0.48~\mu\text{eV}\) and \( h_z=2.5~\text{T} \). The top panel shows the ratio of the Coulomb potential coefficient, $V_0$, compared to the tunneling coupling across the full range of screening, $\lambda$.  The bottom panel shows values of $gS/t$ for fixed $g=0.48~\mu\text{eV}$ as a function of the lattice constant.}  
\end{figure}

\begin{figure}
\includegraphics[width=0.5\textwidth]{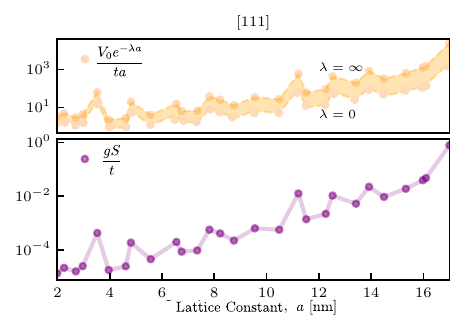}
\centering
\caption{\label{Figure:111}\
\textbf{Effect of lattice constant for [111] orientation.}
The effect of the lattice constant, $a$, on the tunneling coupling, $t$, is plotted for a $[111]$ silicon crystal orientation for fixed values of \( g =0.48~\mu\text{eV}\) and \( h_z=2.5~\text{T} \). The top panel shows the ratio of the Coulomb-potential coefficient, $V_0$, compared to the tunneling coupling across the full range of screening, $\lambda$. The bottom panel shows values of $gS/t$ for fixed $g=0.48~\mu\text{eV}$ as a function of the lattice constant.
}  
\end{figure}

In line with Fig.~\ref{Fig:100}, Figs.~\ref{Figure:111} and \ref{Figure:110} present the calculated tunneling coupling, \( t \), for two of the three orientations of the silicon array, provided for completeness. To plot these figures, we directly use the numerical representation of the wave function and the tunneling coupling as a function of distance from the center of the dopant, provided in Ref.~\cite{gamble2015multivalley}. For more detailed numerical information, please refer to the supplementary material of Ref.~\cite{gamble2015multivalley}.
It is worth highlighting, as demonstrated in Fig.~\ref{Figure:111}, that the $[111]$ silicon orientation provides the most expansive viable region for the emergence of the N\'eel phase. This trait might render the $[111]$ orientation especially intriguing. Nonetheless, a primary concern associated with this orientation is the oscillatory nature of the wave function in the $[111]$ direction, which results in variations in the tunneling coupling.
%

\section{Linear Response Conductance as a Probe}
\label{appendix:linear_conduct}
\noindent
In this section, we review the method for probing the donor systems, as discussed in Sec.~\ref{sec:experimental_probing}, through linear conductance. We begin by modeling the system as an island of electrons hosted by donors interacting with the Fermi sea, using the following Hamiltonian~\cite{le2017extended}:
\begin{align}&H=H^{(\text{lattice})}+H^{(\text{probes})}+H^{(\text{interaction})},
\end{align}
with $H^{(\text{lattice})}$ already defined in Eq.~\eqref{Eq:lattice_model} of the main text and
\begin{align}
    \begin{split}
&H^{(\text{probes})}=\sum_{k \in B^L }\epsilon_k^L c_k^\dagger c_k+\sum_{k \in B^R }\epsilon_k^R c_k^\dagger c_k,
\\
&H^{(\text{interaction})}=\mathcal{C} \sum_{i \in cL,~ j \in L} c_{i}^\dagger c_j +\text{h.c.}
\\
&\hspace{1.65 cm} +\mathcal{C} \sum_{i \in cR,~ j \in R} c_{i}^\dagger c_j +\text{h.c.}
    \end{split}
\end{align}
Here, $L$ and $B^L$ ($R$ and $B^R$) refer to the left (right) probe (in position and momentum space, respectively) with associated momentum modes $\epsilon_k^L$ ($\epsilon_k^R$). $cL$ ($cR$) refers to the sites on a column in the two-dimensional dopant array adjacent to the left (right) probe. The coupling to the left and right probes, $\mathcal{C}$, is assumed equal. This coupling depends on the potential barrier at the probe-dopant interface, and decreases exponentially with the separation between the dopants and the probes~\cite{le2017extended}.

Using the approach outlined in Beenakker's theory of resonant tunneling and linear-response theory for quantum dots~\cite{beenakker1991theory}, one can investigate the linear response conductance of electrons between reservoirs at a specified temperature, $T_{r}$, and a chemical potential, $\mu$, applied to a dopant array with a temperature $T$ perceived as a quantum-dot island. 

We denote by $E_\alpha^{n}$ with $\alpha=0,1,2,\cdots$ the electron energy in level $\alpha$ given $n$ total electrons in the array, and let ${\{n\}}$ be all electron configurations with total electron number $n$ corresponding to the same energy. The linear conductance, $G$, can be derived from the stationary current from the left barrier to the right barrier in the regime linear in the potential difference applied between the two probes, divided by the potential difference. It can be shown to have the form~\cite{beenakker1991theory,le2017extended}:
\begin{equation}
\label{eq:G-relation}
    G=G_{0,T} \sum_{
    \alpha,\alpha'}\sum_{
    \{n\}} 
    Q_{\alpha,\alpha'}^{\{n\}} P^{
    \{n\}}_\alpha[1-f_{\text{FD}}(E_\alpha^
    {n}
    -E_{\alpha'}^{
    n-1}
    \mu)].
\end{equation}
Here, $G_{0,T}=e^2/(k_BT_r)$ and  
\begin{align}
Q^{\{n\}}_{\alpha,\alpha'}=\frac{\Gamma^{L,\{n\}}_{\alpha,\alpha'}\Gamma^{R,\{n\}}_{\alpha,\alpha'}}{\Gamma^{L,\{n\}}_{\alpha,\alpha'}+\Gamma^{R,\{n\}}_{\alpha,\alpha'}}
\end{align}
represents the contribution from the quantum tunneling rates. 
Note that in accordance with Fermi's golden rule, the tunneling rate can be ascertained by observing that the transition of an electron is feasible when its energy matches the energy difference between the state of the array with an additional electron, $E^{n}_\alpha-E^{n-1}_{\alpha'}$. This involves the matrix element of the electrons' creation operator at the edges between the state of the array with $n$ electrons, $\ket{\Psi^{\{n\}}_\alpha}$, and $n-1$ electrons, $\ket{\Psi^{\{n-1\}}_{\alpha'}}$:
\begin{equation}
    \Gamma_{
    \alpha,\alpha'}^{L,\{n\}}=\Gamma \sum_{i \in cL
    }|\langle \Psi^{\{n\}
    }_\alpha|c_i^\dagger |\Psi^{\{n-1\}
    }_{\alpha'}\rangle |^2,
\end{equation}
with $\Gamma=2\pi \mathcal{C}^2$ and $cL$ denoting sites belonging to the leftmost column of the two-dimensional array. A similar expression applies to the right probe. Furthermore, $f_{\text{FD}}(\cdot)$ in Eq.~\eqref{eq:G-relation} is the Fermi-Dirac distribution, $ f_{\text{FD}}(E-\mu)=(1+e^{\frac{E-\mu}{k_BT}})^{-1}$. Finally, the stationary probability function $P^{\{n\}}_\alpha$ in Eq.~\eqref{eq:G-relation} is given by:
\begin{equation}
    P^{\{n\}
    }_\alpha=\frac{e^{-(1/k_BT)(E^{n
    }_\alpha-n\mu)}}{\sum_{
    \alpha'}\sum_{\{n\}} e^{-(1/k_BT)(E^{n
    }_{\alpha'}-n\mu)}}.
\end{equation}
This expression is valid under the assumption that the system and probe possess their thermal state as distinct temperatures. This assumption holds as long as probing occurs at fast time scales compared with the equilibration time of the system-probe composite.

\begin{figure}[t!]
\includegraphics[width=0.5\textwidth]{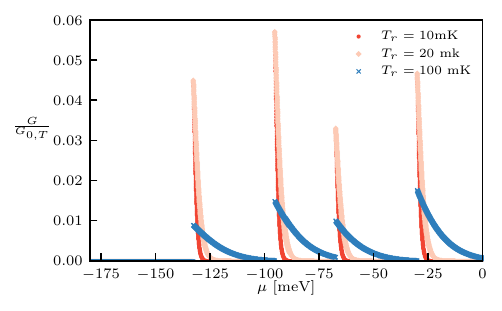}
\centering
\caption{\label{Fig:Conductance_signal_phase2} \textbf{Effect of temperature on linear conductance of a \(2\times 2\) array.} 
The variation in the (normalized) linear-conductance response, \( G \), with respect to chemical potential, \( \mu \), has been simulated for an experimental setup consisting of a \( 2\times 2 \) square dopant array at three distinct reservoir temperatures, \( T_r \), obtained from ED. For the simulation, the following system parameters are used, \( gS/t=4 \times 10^{-5} \), \( h_zS/t=10^{-5} \), and \( h_x/h_z=0.01\).
}  
\end{figure}

To explore the concept of linear conductance as a tool for detecting the different phases of the dopant array, we employed the ED method to simulate the actual wave function of the system, albeit for a limited size. In Fig.~\ref{Fig:Conductance_signal_phase2}, we plot the linear conductance as a function of the chemical potential for a $2 \times 2$ square array over a range of temperatures. We adjust the macroscopic parameters to place the system phase in a nontrivial state. As can be observed in finite-temperature scenarios, with parameters that realistically align with experimental setups around $T_{r}=10 ~\text{mK}$~\cite{wang2022experimental} and $T\ll 1 ~\text{mK}$, the resonance features are both detectable and pronounced. 

\begin{figure}[t!]
\includegraphics[width=0.5\textwidth]{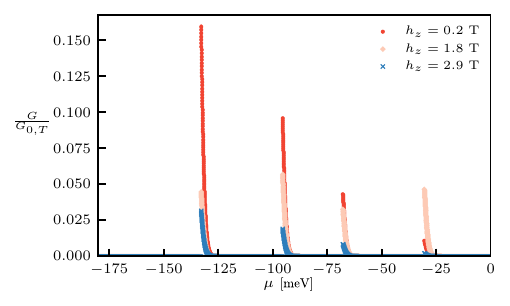}
\centering
\caption{\label{Fig:Conductance_signal_phase}\textbf{Effect of external magnetic field on linear conductance of a \(2\times 2\) array.}
The variation in the (normalized) linear-conductance response, \( G \), with respect to chemical potential, \( \mu \), has been simulated for an experimental setup consisting of a \( 2\times 2 \) square dopant array. 
To investigate the dependency of the system's linear response on its phase, the conductance is simulated using the ED method for three distinct external magnetic fields, \( h_z \), at \( gS/t=4 \times 10^{-5} \). These magnetic fields correspond to different phases of the system, as elaborated in the main text. As we can observe, the conductance profile at $h_z = 1.8$ T is different from the other two, indicating a nontrivial phase.
The variations in these fields lead to noticeable changes in the response profile. The temperature of the array is set to \( T=0.01 \) mK, and the temperature of the reservoir is set to \( T_{r}=10 \) mK.}  
\end{figure}

Moving forward, to investigate the potential for observing phase-transition signatures through conductance behavior, we plot the conductance over a range of external magnetic fields as a macroscopic parameter, for a $2 \times 2$ square array. As illustrated in Fig.~\ref{Fig:Conductance_signal_phase}, the conductance resonance profile varies between nontrivial and trivial phases. Consequently, this suggests that conductance measurement may serve as a probe for exploring and discerning the phases of the system. 

\section{Numerical Simulation Parameters}\label{paramters}
\begin{table*}[h]
\centering
\begin{tabular}{c|c|c|c}
\textbf{Parameter} & \textbf{Description}             & \textbf{Type}    & \textbf{Value}             \\ \hline
$S$                & Nuclear spin                     & Constant         & $1/2$                      \\ \hline
$t$                & Tunneling coupling               & Constant         & $7.5$ meV                  \\ \hline
$g$                & Hyperfine coupling               & Constant         & $0.48~\mu$eV               \\  \hline
$h_z$              & External (longitudinal) magnetic field          & Variable         & [$0-O(1)$] Tesla          \\ \hline
$h_x$              & External (transverse) magnetic field          & Variable         & [$0-O(10^{-2})$] Tesla    \\ \hline
$\beta$            & Inverse of temperature           & Variable         & [$0-O(10^{3})$]            \\ \hline
$V_0$              & Coulomb coupling                 & Constant         & $123 \text{ nm} \cdot \text{meV}$ \\ \hline
$\lambda$          & Coulomb screening                & Variable         & [$0-\infty$] $\text{nm}^{-1}$             \\ \hline
$\mu$              & Chemical potential              & Variable         & [$\pm O(10^{2}))$] meV\\ \hline
$a$                & Lattice constant                 & Constant         & $4.7$ nm                   \\ \hline
$d$                & Distance between pinned spins             & Variable         & [$0-O(N)$]                 \\ \hline
$N$                & Total sites number              & Variable         & [$1-O(10^2)$]              \\ \hline
BC               & Boundary condition              & Constant         & Periodic             \\
\end{tabular}
\caption{Summary of numerical values of parameters and other features adopted in the simulations of this work.}
\label{tab:simulation_parameters}
\end{table*}
\noindent
In the numerical simulations conducted for this study, specific values of parameters within experimentally feasible ranges have been used. Here, we provide a list of these parameters to ensure completeness and reproducibility. 

\end{document}